\documentclass[12pt,eadjoint tfnpsf]{article}

\catcode`\@=11
\@addtoreset{equation}{section}

\global\arraycolsep=1pt

\usepackage{amsbsy,amssymb,latexsym,amsfonts, amsmath}
\usepackage{mathrsfs}
\usepackage{graphicx}
\usepackage{youngtab}

\RequirePackage[dvips,usenames]{color}
\definecolor{fireblick}{rgb}{0.698039,0.133333,0.133333}

\newcommand{\beq}{\begin{equation}}
\newcommand{\eeq}{\end{equation}}
\newcommand{\bea}{\begin{eqnarray}}
\newcommand{\eea}{\end{eqnarray}}

\renewcommand{\thefootnote}{\fnsymbol{footnote}}

\begin{document}
%
%
\begin{titlepage}

\begin{flushright}
\normalsize
~~~~
SISSA  34/2013/MATE-FISI\\
\end{flushright}


\vspace{80pt}

\begin{center}
{\LARGE 
Vortex partition functions, wall crossing 
\\ 
and 
\\ 
\vspace{.5cm}
equivariant Gromov-Witten invariants
}
\end{center}

\vspace{25pt}

\begin{center}
{
Giulio Bonelli$^{\heartsuit\spadesuit}$, Antonio Sciarappa$^{\heartsuit}$, Alessandro Tanzini$^{\heartsuit}$ and Petr Vasko$^{\heartsuit}$
}\\
%
\vspace{15pt}
%
$^{\heartsuit}$
International School of Advanced Studies (SISSA) \\via Bonomea 265, 34136 Trieste, Italy 
and INFN, Sezione di Trieste \\
\vspace{15pt}
$^{\spadesuit}$
I.C.T.P.\\ Strada Costiera 11, 34014 Trieste, Italy
\end{center}
%
\vspace{20pt}
%

In this paper we identify 
the problem of equivariant vortex counting 
in a $(2,2)$ supersymmetric two dimensional quiver gauged linear sigma model
with 
that of computing the equivariant Gromov-Witten invariants of the GIT quotient target space
determined by the quiver.
We provide new contour integral formulae for the ${\cal I}$ and ${\cal J}$-functions
encoding the equivariant quantum cohomology of the target space. Its chamber structure
is shown to be encoded in the analytical properties of the integrand.
This is explained both via general arguments and by checking several key cases.
We show how several results in equivariant Gromov-Witten theory follow just by deforming 
the integration contour. 
In particular we apply our formalism to 
compute Gromov-Witten
invariants of the
$\mathbb{C}^3/\mathbb{Z}_n$ orbifold, of the Uhlembeck (partial) compactification
of the moduli space of instantons on $\mathbb {C}^2$ and of $A_n$ and $D_n$ singularities
both in the orbifold and resolved phases.
Moreover, we 
analyse
dualities of quantum cohomology rings 
of holomorphic vector bundles over Grassmannians,
which are relevant to BPS Wilson loop algebrae.



\vfill

\setcounter{footnote}{0}
\renewcommand{\thefootnote}{\arabic{footnote}}

\end{titlepage}

\section{Introduction}
\label{sec:intro}
One of the most exciting aspects of supersymmetric quantum field theories is the possibility to get exact non perturbative solutions via 
a variety of techniques. In this paper we will focus on two dimensional 
gauge theories with four supersymmetries.
In these cases the non perturbative aspects are captured by vortex counting.
This was initially developed in \cite{2007JHEP...08..052S} who applied the equivariant localization of \cite{2002hep.th....6161N} to two dimensional gauge theories giving explicit vortex partition function formulas, which recently attracted attention in the context of AGT correspondence 
\cite{Alday:2009aq} and knot theory \cite{Dimofte:2010tz}.
Vortex partition functions have been related to CFT degenerate conformal blocks and to topological strings in
\cite{Dimofte:2010tz}\cite{Bonelli:2011fq}\cite{Bonelli:2011wx}\cite{Kozcaz:2010yp}\cite{Kanno:2011fw}\cite{Bulycheva:2012ct}.
General contour integral formulae for vortex counting have been obtained in \cite{Benini:2012ui}
\cite{Doroud:2012xw}
in the study of supersymmetric partition functions on $S^2$.
These partition functions have been conjectured to compute the quantum K\"ahler potential of the target space of the corresponding infrared NLSM
in \cite{Jockers:2012dk}. Evidence of this conjecture was provided in \cite{Gomis:2012wy}.
Further studies along these lines have been presented in 
\cite{Park:2012nn}\cite{Sharpe:2012ji}\cite{2013JHEP...05..102H}\cite{Halverson:2013eua}\cite{Sharpe:2013bwa}.
In this paper we will elaborate on these issues from a different viewpoint by using supersymmetric localization on $S^2$
to provide new contour integral formulae for the 
${\cal I}$ and ${\cal J}$-functions describing the equivariant quantum cohomology of GIT quotients in terms of Givental's formalism
\cite{1996alg.geom..3021G}
and its extension to non abelian quotients in terms of quasi-maps \cite{2011arXiv1106.3724C}.

One of the implications of our results is thus that the equivariant vortex partition functions 
contain not only information about the Gromov-Witten invariants of the IR target space, but also
their gravitational descendants. As will be explained more in detail in Sec.2, this is a consequence
of the equivariant localization procedure with respect to a supersymmetric charge that closes
on $U(1)_R$ rotations of the sphere. From the geometrical viewpoint one thus considers 
$S^1$-equivariant maps from a sphere with marked North and South pole, where the gravitational descendants are inserted, to the target space.

We provide general rules for the calculation of supersymmetric spherical partition functions of quiver gauge theories
and the corresponding ${\cal I}$-functions. 
Our formalism applies to both compact and non compact K\"ahler manifolds with $c_1 \geq 0$.
One key result that we will obtain is the possibility of analyzing the chamber structure and wall-crossings of the GIT quotient 
moduli space in terms of integration contour choices. 
In particular, we will obtain explicit description of the equivariant quantum cohomology and
chamber structure for the resolutions of ${\mathbb C}^3/ {\mathbb Z}_n$ orbifolds and for the Uhlembeck partial compactification of the instanton moduli space.

We remark that, as observed in \cite{2013arXiv1302.2164K}, the OPE algebra of circular BPS Wilson loops in three dimensional supersymmetric gauge theories can be reduced in some cases to the equivariant quantum K-ring of certain quasi projective varieties. In particular this led to conjecture
an equivalence of the quantum cohomology rings of suitable vector bundles over complex Grassmannians using 3d dualities and circle compactification.
We will use our methods to prove this conjecture in Section 4.


The paper is organized as follows.
In Section 2 we provide a general discussion about the relation between the spherical partition function 
of a given GLSM and the quantum cohomology of the space it flows to in the IR in terms of ${\cal I}$ and 
${\cal J}$-functions.
In Section 3 and 4 we provide several examples of calculations of the quantum cohomology of 
abelian and non-abelian GIT quotients.
We study in particular the chamber structure 
of the crepant resolution
of the orbifold ${\mathbb C}^3/{\mathbb Z}_n$ in subsection 3.4.2
and 
of the ADHM moduli space in subsection 4.4.
The duality between Grassmannians is discussed in subsection 4.1 (with details in the Appendix)
and quiver gauge theories are discussed in subsection 4.2 and 4.3.
Finally, in Section 5 we draw our conclusions and discuss further directions.

\section{Gauge Linear Sigma Models, stability conditions and wall crossing}

In this section we discuss 
how the exact equivariant partition functions 
of general ${\cal N}=(2,2)$ gauged linear sigma models 
on the two-sphere with a $U(1)$ vector $R$-symmetry
\cite{Benini:2012ui,Doroud:2012xw}
encode the quantum cohomology of the target IR geometry
in various stability chambers
and 
the wall crossing among them.

The partition function for a given gauge 
group\footnote{The localization applies to any classical Lie group ABCDEFG. In this paper we will focus on the $U(N)$ case.}
$G$ and matter in the representation $R$
depends on the twisted masses which can be coupled to the system breaking its continuous
flavor symmetry group $G_F$ to its maximal abelian subgroup $T_F$. The theory in general allows
a gauge invariant holomorphic non singular superpotential ${\cal W}$.

The resulting object, in the Coulomb branch localization scheme, 
is given as an integral over the Cartan algebra $t_G$ of the gauge symmetry group
\beq
Z^{S^2}=\frac{1}{|W(G)|}\sum_{\vec m\in{\mathbb Z}^{r_G}}
\int_{t_G} d{\vec \tau} e^{-S_{cl}} \mu_{G} \mu_{R}
\label{cmtpr}
\eeq
where
$|W(G)|$ is the order of the Weyl group of $G$,
$r_G={\rm dim}t_G$ is the rank of the gauge group.
$S_{cl}=-4\pi\vec \xi\cdot{\vec \tau}+i\vec \theta\cdot \vec m$
is the classical action of the GLSM depending on the 
FI parameters vector $\vec \xi$ (one for each $U(1)$ factor in $G$), 
the magnetic fluxes $\vec m$ and the theta-angles $\vec \theta$.
More specific rules for quiver gauge theories will be presented in Section 4.

In \eqref{cmtpr}
$\mu_G$ is the one loop determinant of the gauge multiplet
\beq
\prod_{r<s}^{r_G}\left(\dfrac{m_{rs}^{2}}{4}-\tau_{rs}^{2}\right),
\eeq
where $m_{rs}=m_r-m_s$ and $\tau_{rs}=\tau_r-\tau_s$,
and 
$\mu_R$ is the one-loop determinant of the matter multiplets
\beq
\prod_{\rho\in R}\frac{\Gamma\left({\bf q}/2+r\rho(\tau)-\frac{\rho(m)}{2}\right)}
{\Gamma\left(1- {\bf q}/2-r\rho(\tau)-\frac{\rho(m)}{2}\right)}
\eeq
where ${\bf q}$ is the vector $R$-charge, $r$ is the radius of $S^2$ and 
$\rho$ is the weight of the representation the matter multiplet belongs to.

Thanks to \eqref{cmtpr}, the computation of the partition function is reduced to residues evaluation as 
\begin{equation}
\oint \prod_{r=1}^{r_G} \dfrac{d (r \lambda_{r})}{2\pi i} (z \bar{z})^{-r \lambda_{r}} 
Z_{\text{1l}} Z_{\text{v}} Z_{\text{av}} \label{int}
\end{equation}
where $z=e^{-2\pi\vec\xi+i\vec\theta}$ labels the different vortex sectors, $(z \bar{z})^{-r \lambda_{r}}$ is
a contribution from 
the classical action, $Z_{\text{v}}$ is the equivariant vortex partition function
on the north pole patch, $Z_{\text{av}}$ is the equivariant vortex partition function
on the south pole patch
and $Z_{\text{1l}}$ is the remnant one-loop measure.
The contour of integration in \eqref{int} crucially depends on the choice of 
the FI-parameters and this, as we will specify better in a moment, encodes the geometric interpretation 
of the partition function.

One can actually read the GLSM data from a geometric perspective as in the following table \cite{Witten:1993yc}.
\begin{table}[h!]
\begin{center}
\begin{tabular}{c|c}
GLSM             & GW                            \\ \hline
matter fields    & quasi-affine variety {\cal A} \\ \hline
gauge group $G$  & $G_{\mathbb C}$ action on {\cal A}      \\ \hline
F/D-terms        & stable GIT quotient ${\cal A}//G_{\mathbb C}$ \\ \hline
\end{tabular} 
\caption{GLSM vs. GIT quotient}
\end{center}
\end{table} 
Let us remark that the GLSM counterpart of the GIT stability condition is in the D-term equation 
which crucially depends on the FI parameters. The different stability chambers are 
in one-to-one correspondence with the phases of the GLSM as defined by the domains 
of the FI parameters. As far as the models that we study in this paper are concerned, for Abelian quotients, when the FIs are large and positive
one describes a geometric phase, namely a NLSM on a K\"ahler target manifold
\cite{Jockers:2012dk},
while for negative FIs the GLSM is in a Landau-Ginsburg phase describing an orbifold target space.
In the non-Abelian case, the possibility of having a reflection symmetry on the FI opens up
leaving the orbifold phases at the fixed point of the reflection.

From the perspective of Eq. \eqref{int}, different FI phases imply that the integral converges 
at different asymptotic regions of the $\tau$-plane imposing different
choices of the contour integral. As we will largely exemplify in the following this allows to describe 
the quantum cohomology of the corresponding GIT quotients in the different stability chambers.
In particular, we will study the crepant resolution conjecture 
for both abelian and non-abelian quotients, focusing on 
${\mathbb C}^3/{\mathbb Z}_n$ and on the Uhlembeck (partial) compactification of the ADHM moduli space respectively.
This provides conjectural formulas for the ${\cal I}$ and ${\cal J}$-functions which are shown to reduce
in the relevant particular cases to those of 
\cite{2006math......8481C} for the $\mathbb{Z}_3$ and $\mathbb{Z}_4$ orbifolds
and of \cite{2006math.....10129B} for the symmetric product of points in $\mathbb{C}^2$
(see later sections).

Let us now provide more details on how the quantum cohomology of the target GIT
quotients is computed from the spherical partition function. 
It has been argued in \cite{Jockers:2012dk}  that the spherical partition
function computes the vacuum amplitude of the NLSM in the infrared 
\beq
\langle\bar 0 | 0 \rangle= e^{-K}
\label{00}
\eeq
where $K$ is the quantum K\"ahler potential of the target space $X$. A general argument for
the validity of this conjecture has been provided in  \cite{Gomis:2012wy}, whose main idea
goes as follows.
One considers the spherical partition function on the squashed two-sphere
discovering that it is independent on the squashing parameter.
Then the limit of extreme squashing is identified with the 
topological-antitopological  fusion
$\langle\bar 0 | 0 \rangle$.
We remark that although \cite{Gomis:2012wy} focused on Calabi-Yau target manifolds
their arguments apply also to Fano manifolds, for which both the A and B-twist are well defined,
the latter being a Landau-Ginzburg model with cylinder as its target space. Indeed we will discuss
several examples of this type including (weighted) projective spaces and (partial) flag manifolds.

Let us now draw some further steps in the analysis of the spherical partition function
from a general viewpoint.
Let us rewrite the above vacuum amplitude
in a way which is more suitable for our purposes.
Following \cite{Bershadsky:1993ta,Dubrovin:1994hc}, let us introduce the flat sections $V_a$ 
of the Gauss-Manin connection
spanning the vacuum bundle
of the theory and
satisfying
\beq
\left(\hbar D_a \delta_b^c + C_{ab}^c\right)V_c=0.
\label{1}\eeq
where $D_a$ is the covariant derivative on the vacuum line bundle and $C_{ab}^c$ are the
coefficients of the OPE in the chiral ring of observables $\phi_a\phi_b =C_{ab}^c \phi_c$. 
The observables $\{\phi_a\}$ provide a basis for the vector space
of chiral ring operators $H^0(X)\oplus H^2(X)$ with $a=0,1,\ldots,b^2(X)$, 
$\phi_0$ being the identity operator.
The parameter $\hbar$ is the spectral parameter of the Gauss-Manin connection. 
Specifying the case $b=0$ in (\ref{1}), we find that 
$
V_a=-\hbar D_aV_0
$
which means that the flat sections are all generated by the fundamental solution ${\cal J}:=V_0$
of the equation
\beq
\left(\hbar D_a D_b + C_{ab}^c D_c\right){\cal J}=0
\label{2}\eeq
In order to uniquely fix the solution to (\ref{2}) one needs to supplement some further information 
about the dependence on the spectral parameter. This is usually done by 
combining the dimensional analysis of the theory with the the $\hbar$ dependence
by fixing
\beq
\left(
\hbar\partial_{\hbar}+{\cal E}
\right){\cal J}=0
\label{3}\eeq
where the covariantly constant Euler vector field ${\cal E}=\delta^aD_a$,  $\delta^a$ being the vector of scaling dimensions of the coupling constants, scales with weight one the 
chiral ring structure constants as ${\cal E} C_{ab}^c=C_{ab}^c$ to ensure compatibility between 
(\ref{2}) and (\ref{3}).

The metric on the vacuum bundle is given by a symplectic pairing of the flat sections
$g_{\bar a b}= \langle \bar a|b\rangle = V_{\bar a}^t E V_b$
and in particular
the vacuum-vacuum amplitude, that is the
the spherical partition function, can be written as the symplectic pairing
\beq
\langle\bar 0 | 0 \rangle = {\cal J}^t E {\cal J}
\label{vac}
\eeq
for a suitable symplectic form $E$ \cite{Bershadsky:1993ta} that will be specified later.

Let us remark that in the case of non compact target, the Quantum Field Theory has to be studied in 
the equivariant sense to regulate its volume divergences already visible in the constant map contribution.
This is accomplished by turning on the relevant twisted masses for matter fields.
From the mathematical viewpoint, this amounts to work in the context of equivariant cohomology of 
the target space $H^\bullet_T(X)$ where $T$ is the torus acting on $X$.
The values of the twisted masses assign the weights of the torus action. 

We point out that there is a natural correspondence of the results of supersymmetric localization on the two-sphere
with the formalism developed by Givental for the computation of the flat section ${\cal J}$.
Indeed
the computation of the spherical partition function 
makes use of a supersymmetric charge which closes on a $U(1)$ isometry
of the sphere, whose fixed points are the north and south pole. From the string viewpoint it therefore
describes the embedding in the target space of a spherical world-sheet {\it with two marked points}
where the gravitational descendant are inserted. 
This is precisely the setting of $S^1$-equivariant Gromov-Witten invariants considered by Givental in \cite{1996alg.geom..3021G}  by studying
equivariant
holomorphic maps with respect to the maximal torus of the sphere automorphisms 
$S^1\subset PSL(2,\mathbb{C})$. This is identified with the $U(1)$ isometry to which
the supersymmetry algebra squares. As an important consequence,  
the equivariant parameter $\hbar$ of Givental's $S^1$ action gets identified with the one of the 
vortex partition functions arising in the localization of the spherical partition function.
An excellent review of Givental's formalism can be found in \cite{2001math.....10142C}, here we will highlight 
the aspects that are strictly relevant for the subsequent discussions. 
The ${\cal J}$-function can be computed from a set of oscillatory integrals, the so called ``${\cal I}$-functions''
which are generating functions of hypergeometric type in the variables $\hbar$ and $Q_i$, 
where $Q_i = e^{- t^i}$, $t^i$ being the complexified K\"ahler parameters and $i=1,\ldots,b_2(X)$.
We observe that Givental's formalism has been developed originally for abelian quotients, more precisely for complete intersections in quasi-projective toric varieties.
In this case, the ${\cal I}$ function is the generating function of solutions of the Picard-Fuchs equations for the mirror manifold $\check X$ of $X$ 
and as such can be expressed in terms of periods on $\check X$.
From the viewpoint of the spherical partition function this has also a very nice
direct interpretation by an alternative rewriting of the vacuum amplitude \eqref{vac}.
Indeed, by mirror symmetry one can rewrite, in the Calabi-Yau case
\beq
\langle\bar 0 | 0 \rangle = i \int_{\check X} \overline\Omega\wedge\Omega = \Pi^t S \Pi
\label{pispi}
\eeq
where $\Pi = \int_{\Gamma^i} \Omega $ is the period vector and $S$ is the symplectic pairing.
The components of the ${\cal I}$-function can be identified with the components of the period vector $\Pi$.
More in general one can consider
 an elaboration of the integral form of the 
spherical partition function worked out in \cite{Gomis:2012wy}, where the integrand is rewritten in
a mirror symmetric manifest form, by expressing the ratios of $\Gamma$-functions appearing 
in the Coulomb branch representation as
\beq
\frac{\Gamma(\Sigma)}{\Gamma(1-\bar\Sigma)}=\int_{Im(Y)\sim Im(Y)+2\pi} \frac{d^2 Y}{2\pi i} e^{\left[e^{-Y}-\Sigma Y - c.c.\right]}
\eeq
to obtain the right-hand-side (\ref{pispi}) and then by applying the Riemann bilinear identity, one gets the left-hand side.
The resulting integrals, after the integration over the Coulomb parameters and  
independently on the fact that the mirror representation is geometric or not,
are then of the oscillatory type
\beq
\Pi_i=\oint_{\Gamma_i} d\vec{Y} e^{r{\cal W}_{eff}\left(\vec{Y}\right)}
\label{osc}
\eeq
where the effective variables $\vec{Y}$ and potential ${\cal W}_{eff}$
are the remnants parametrizing the constraints imposed by the integration over the Coulomb parameters
before getting to (\ref{osc}). Eq.(\ref{osc}) is also the integral representation of Givental's ${\cal I}$-function

for general Fano manifolds \cite{2001math.....10142C}. 
Non-abelian quotients have been studied in \cite{2011arXiv1106.3724C} in terms of quasi-maps theory which is the mathematical
counterpart of the GLSM.

Let us now state the dictionary between Givental's formalism and the spherical partition function 
\beq
{Z}^{S^2} = \oint 
d\lambda Z_{\text{1l}} \left(z^{-r|\lambda|} Z_{\rm v}\right) \left(\bar z^{-r|\lambda|} Z_{\rm av}\right)
\label{ciao} 
\eeq
with $d\lambda=\prod_{\alpha=1}^{\rm rank}d\lambda_\alpha$ and $|\lambda|=\sum_\alpha \lambda_\alpha$.
Our claim \cite{Bonelli:2013rja} is that $Z_{\rm v}$ is the ${\cal I}$-function of the target space $X$  
upon identifying the vortex counting parameter $z$ with $Q$, $\lambda_\alpha$ with the generators of the
equivariant cohomology and $r=1/\hbar$. 
More precisely, the chamber structure of the GIT quotient is encoded in the choice of the FI parameters
and the subsequent choice of integration contours. 
In particular, in the geometric phase with all the FIs large and positive, 
the vortex counting parameters are identified with the exponentiated complex K\"ahler parameters, 
while, in the orbifold phase
they label the twisted sectors of the orbifold itself or, in other words, the basis of orbifold cohomology.

The ${\cal J}$-function -- needed to 
compute the equivariant Gromov-Witten invariants of $X$ -- is then obtained from the ${\cal I}$-function after a suitable 
normalisation procedure which has been described in \cite{Bonelli:2013rja}.
Actually, in some cases one can show that 
the ${\cal I}$ and the ${\cal J}$-functions coincide and that
this normalisation procedure is not required. 
This is the case of Fano manifolds and ADHM moduli space for rank higher than one.

A further normalization is then required for the one-loop term
in order to reproduce the classical intersection cohomology on the target manifold.
In this normalization, 
the spherical partition function coincides with the symplectic pairing 
\eqref{vac}
and in particular
the one-loop part reproduces in the $r\to0$ limit the (equivariant) volume
of the target space.

The above conjecture will be checked for several abelian and non abelian GIT quotients 
in the subsequent sections.

\section{Abelian GLSMs}

\subsection{Projective spaces}

Let us start with the basic example, that is $\mathbb{P}^{n-1}$. Its sigma model matter content consists of $n$ chiral fields of charge $1$ with respect to the $U(1)$ gauge group. In general, the Fayet-Iliopoulos parameter runs \cite{Doroud:2012xw}; in our case 
\begin{equation}
\xi_{\text{ren}} = \xi - \frac{n}{2 \pi} \log (r M)
\end{equation}
with $M$ a SUSY-invariant ultraviolet cut-off. Notice that in the Calabi-Yau case the sum of the charges is zero, therefore\footnote{We will also assume that $\theta_{\text{ren}} = \theta + (s-1)\pi$, with $s$ rank of the gauge group; this implies $\theta_{\text{ren}} = \theta$ for abelian gauge groups. This is necessary in order to reproduce the known results in the mathematical literature for Grassmannians, flag manifolds, and the Hilbert scheme of points; this shift should come from integrating out the $W$ bosons, but we do not have a detailed explanation for it.}
$\xi_{\text{ren}} = \xi$.
\\
By defining\footnote{We are following the notation of \cite{Benini:2012ui}, but we work with dimensionless partition functions: this means that in our integrals it appears $d(r \sigma)$ instead of $d \sigma$.} $\tau = -i r \sigma $ the $\mathbb{P}^{n-1}$ partition function reads  
\begin{equation}
Z_{\mathbb{P}^{n-1}}=\sum_{m\in\mathbb{Z}}\int \frac{\mathrm{d}\tau}{2\pi i}e^{4\pi \xi_{\text{ren}} \tau - i \theta_{\text{ren}} m} \left( \frac{\Gamma\left(\tau -\frac{m}{2}\right)}{\Gamma\left(1-\tau -\frac{m}{2}\right)}\right)^n \label{cpn}
\end{equation}
With the change of variables 
\begin{equation}
\tau = -k + \frac{m}{2} + r M \lambda \label{poles}
\end{equation}
we are resumming over all the poles, which are at $\lambda = 0$. Equation \eqref{cpn} then becomes
\begin{equation}
Z_{\mathbb{P}^{n-1}} = \oint \dfrac{d (r M \lambda)}{2\pi i} Z_{\text{1l}}^{\mathbb{P}^{n-1}}Z_{\text{v}}^{\mathbb{P}^{n-1}}Z_{\text{av}}^{\mathbb{P}^{n-1}}
\end{equation}
where $z=e^{-2 \pi \xi + i \theta}$ and
\begin{equation}
\begin{split}
Z_{\text{1l}}^{\mathbb{P}^{n-1}} = & \, (r M)^{-2nr M \lambda} \left( \frac{\Gamma (rM \lambda)}{\Gamma(1-rM \lambda)}\right)^n \\
Z_{\text{v}}^{\mathbb{P}^{n-1}} = & \, z^{-r M \lambda}\sum_{l\geq 0} \dfrac{[(rM)^n z]^l}{(1-rM \lambda)_l^n} \\
Z_{\text{av}}^{\mathbb{P}^{n-1}} = & \, \bar{z}^{-r M \lambda}\sum_{k\geq 0} \dfrac{[(-rM)^n\bar{z}]^k}{(1-rM \lambda)_k^n}\\
\end{split}
\end{equation}
The Pochhammer symbol $(a)_k$ is defined as
\begin{equation}
(a)_k = \left\{ 
\begin{array}{cc}
\prod_{i=0}^{k-1} (a+i) & \,\,\text{for}\,\, k>0\\
1 & \,\,\text{for}\,\, k=0\\
\prod_{i=1}^{-k} \dfrac{1}{a-i} & \,\,\text{for}\,\, k<0
\end{array}
\right. \label{poc}
\end{equation}
The $\mathcal{I}$-function is given by $Z_{\text{v}}^{\mathbb{P}^{n-1}}$, and coincides with the one given in the mathematical literature\footnote{This was already observed in this particular case 
in \cite{Dimofte:2010tz}.},
\begin{equation}
\mathcal{I}_{\mathbb{P}^{n-1}} (H, \hbar ; t) = e^{\frac{t H}{\hbar}} \sum_{d \geq 0} \dfrac{[(\hbar)^{-n} e^t]^d}{(1+H/\hbar)_d^n}
\end{equation}
if we identify $\hbar = \frac{1}{rM},\, H = -\lambda,\, t = \ln z$. 
The antivortex contribution is the conjugate ${\cal I}$-function, with 
$\hbar = -\frac{1}{rM},\, H = \lambda $ and $\bar{t} = \ln \bar{z}$. 
The hyperplane class $H$ satisfies $H^n = 0$; in some sense the integration variable $\lambda$ satisfies the same relation, because the process of integration will take into account only terms up to $\lambda^{n-1}$ in $Z_{\text{v}}$ and $Z_{\text{av}}$. \\

Complete intersections in $\mathbb{P}^{n-1}$ of type $(q_0, \ldots, q_m)$, $q_j > 0$ can be obtained by adding chiral fields of charge $(-q_0, \ldots, -q_m)$. This means that the integrand in \eqref{cpn} gets multiplied by
\begin{equation}
\prod_{j=0}^m \frac{\Gamma\left(\frac{R_j}{2}-q_j \tau + q_j \frac{m}{2}\right)}{\Gamma\left(1-\frac{R_j}{2} + q_j \tau + q_j \frac{m}{2}\right)}
\end{equation}
The poles are still as in \eqref{poles}, but now
\begin{equation}
\begin{split}
Z_{\text{1l}}^{\mathbb{P}^{n-1}} = & \, (r M)^{-2r M (n- \vert q \vert) \lambda} \left( \frac{\Gamma (rM \lambda)}{\Gamma(1-rM \lambda)}\right)^n \prod_{j=0}^m \frac{\Gamma\left(\frac{R_j}{2}-q_j rM \lambda\right)}{\Gamma\left(1-\frac{R_j}{2} + q_j rM \lambda \right)}\\
Z_{\text{v}}^{\mathbb{P}^{n-1}} = & \, z^{-r M \lambda}\sum_{l\geq 0}(-1)^{\vert q \vert l} [(rM)^{n-\vert q \vert } z]^l \dfrac{\prod_{j=0}^m (\frac{R_j}{2}-q_j r M \lambda)_{q_j l}}{(1-rM \lambda)_l^n} \\
Z_{\text{av}}^{\mathbb{P}^{n-1}} = & \, \bar{z}^{-r M \lambda}\sum_{k\geq 0}(-1)^{\vert q \vert k} [(-rM)^{n - \vert q \vert }\bar{z}]^k \dfrac{\prod_{j=0}^m (\frac{R_j}{2}-q_j r M \lambda)_{q_j k}}{(1-rM \lambda)_k^n} \\
\end{split}
\end{equation}
where $\vert q \vert = \sum_{j=0}^n q_j $ and $R_j$ is the $R$-charge of the $j$-th field. Notice that, if we want to describe a bundle over a space, we should set $R_j = 0$ and add twisted masses in the contributions coming from the fibers, since we want to separate the different cohomology generators (i.e. the different integration variables); we will do this explicitly when needed. On the other hand, complete intersections do not require and do not allow twisted masses, because the insertion of the superpotential breaks all flavour symmetry; moreover, since the superpotential must have $R$-charge $2$, we will need some $R_j \neq 0$ (see the example of the quintic below).

\subsubsection{Equivariant projective spaces}

The same computation can be repeated in the more general equivariant case, with twisted masses turned on. In this case, the partition function reads (rescaling the twisted masses as $a_i \rightarrow M a_i$ in order to have dimensionless parameters)
\begin{equation}
Z_{\mathbb{P}^{n-1}}^{\text{eq}}=\sum_{m\in\mathbb{Z}}\int \frac{\mathrm{d}\tau}{2\pi i}e^{4\pi \xi_{\text{ren}} \tau - i \theta_{\text{ren}} m} \prod_{i=1}^n \frac{\Gamma\left(\tau -\frac{m}{2} + i r M a_i \right)}{\Gamma\left(1-\tau -\frac{m}{2} - i r M a_i \right)} 
\end{equation}
Choosing poles at
\begin{equation}
\tau = -k + \frac{m}{2} - i r M a_j + r M \lambda 
\end{equation}
we arrive at
\begin{equation}
Z_{\mathbb{P}^{n-1}}^{\text{eq}} = \sum_{j=1}^n \oint \dfrac{d (r M \lambda)}{2\pi i} Z_{\text{1l, eq}}^{\mathbb{P}^{n-1}}Z_{\text{v, eq}}^{\mathbb{P}^{n-1}}Z_{\text{av, eq}}^{\mathbb{P}^{n-1}}
\end{equation}
where 
\begin{equation}
\begin{split}
Z_{\text{1l, eq}}^{\mathbb{P}^{n-1}} = & \, (z \bar{z})^{i r M a_j} (r M)^{-2nr M \lambda} \prod_{i=1}^n \frac{\Gamma (rM \lambda + i r M a_{ij})}{\Gamma(1-rM \lambda - i r M a_{ij})}  \\
Z_{\text{v, eq}}^{\mathbb{P}^{n-1}} = & \, z^{-r M \lambda}\sum_{l\geq 0} \dfrac{[(rM)^n z]^l}{\prod_{i=1}^n (1-rM \lambda - i r M a_{ij})_l} \\
Z_{\text{av, eq}}^{\mathbb{P}^{n-1}} = & \, \bar{z}^{-r M \lambda}\sum_{k\geq 0} \dfrac{[(-rM)^n\bar{z}]^k}{\prod_{i=1}^n(1-r M \lambda - i r M a_{ij})_k} \\
\end{split}
\end{equation}
and $a_{ij} = a_i-a_j$. Since there are just simple poles, the integration can be easily performed:
\begin{equation}
\begin{split}
Z_{\mathbb{P}^{n-1}}^{\text{eq}} = & \, \sum_{j=1}^n (z \bar{z})^{i r M a_j} \prod_{i\neq j =1}^n \frac{1}{i r M a_{ij}} \frac{\Gamma (1 + i r M a_{ij})}{\Gamma(1 - i r M a_{ij})} \\
& \, \sum_{l\geq 0} \dfrac{[(rM)^n z]^l}{\prod_{i=1}^n (1 - i r M a_{ij})_l} \sum_{k\geq 0} \dfrac{[(-rM)^n\bar{z}]^k}{\prod_{i=1}^n(1 - i r M a_{ij})_k} \\
\label{abcd}
\end{split}
\end{equation}
In the limit $rM\to0$ the one-loop contribution (see the first line of \eqref{abcd})
provides the equivariant volume of the target space:
\begin{equation}
\text{Vol}(\mathbb{P}^{n-1}_{\text{eq}}) = \sum_{j=1}^n (z \bar{z})^{i r M a_j} \prod_{i\neq j =1}^n \frac{1}{i r M a_{ij}} = \sum_{j=1}^n e^{-4 \pi i \xi r M a_j} \prod_{i\neq j =1}^n \frac{1}{i r M a_{ij}}
\end{equation}
Using the fact that
\begin{equation}
\lim_{r \to 0} \,\,  \sum_{j=1}^n \dfrac{e^{-4 \pi i \xi r M a_j}}{(4 \xi)^{n-1}}  \prod_{i\neq j =1}^n \frac{1}{i r M a_{ij}} = \dfrac{\pi^{n-1}}{(n-1)!}
\end{equation}
we find the non-equivariant volume
\begin{equation}
\text{Vol}(\mathbb{P}^{n-1}) = \dfrac{(4 \pi \xi)^{n-1}}{(n-1)!}
\end{equation}

\subsubsection{Weighted projective spaces}

\noindent Another generalization consists in studying the weighted projective space $\mathbb{P}^{\textbf{w}} = \mathbb{P}(w_0, \ldots, w_n)$, which has been studied from the mathematical point of view in \cite{2006math......8481C}. This can be obtained by considering an $U(1)$ gauge theory with $n+1$ fundamentals of (positive) integer charges $w_0, \ldots, w_n$. The partition function reads
\begin{equation}
Z \,=\, \sum_m \int \dfrac{d \tau}{2 \pi i} e^{4 \pi \xi_{\text{ren}} \tau - i \theta_{\text{ren}} m} \prod_{i = 0}^{n} \dfrac{\Gamma(w_i \tau - w_i \frac{m}{2})}{\Gamma(1 - w_i \tau - w_i \frac{m}{2})} \label{wpspa}
\end{equation}
so one would expect $n+1$ towers of poles at
\begin{equation}
\tau \,=\, \frac{m}{2} - \frac{k}{w_i} + r M \lambda \,\,\,,\,\,\, i=0\,\ldots \, n
\end{equation}
with integration around $r M \lambda = 0$. Actually, in this way we might be overcounting some poles if the $w_i$ are not relatively prime, and in any case the pole $\tau = 0$ is always counted $n+1$ times. In order to solve these problems, we will set
\begin{equation}
\tau \,=\, \frac{m}{2} - k + r M \lambda - F 
\end{equation}
where $F$ is a set of rational numbers defined as
\begin{equation}
F = \big\{ \,\frac{d}{w_i} \,\,\,/\,\,\, 0 \leq d < w_i \, , \,\,\, d \in \mathbb{N} \,,\,\,\, 0 \leq i \leq n \, \big\}
\label{effe}\end{equation}
and every number has to be counted only once. Let us explain this better with an example: if we consider just $w_0 = 2$ and $w_1 = 3$, we find the numbers $(0, 1/2)$ and $(0, 1/3, 2/3)$, which means $F = (0, 1/3, 1/2, 2/3)$; the multiplicity of these numbers reflects the order of the pole in the integrand, so we will have a 
double pole (counted by the double multiplicity of $d=0$) and three simple poles.\\
\noindent The partition function then becomes
\begin{equation}
Z \,=\, \sum_F \oint \dfrac{d(r M \lambda)}{2 \pi i} Z_{\text{1l}} \, Z_{\text{v}} \, Z_{\text{av}}
\end{equation}
with integration around $r M \lambda = 0$ and
\begin{equation}
\begin{split}
Z_{\text{1l}} = & \,(rM)^{- 2 \vert w \vert r M \lambda - 2 \sum_{i=0}^n (\omega[w_i F] - \langle w_i F \rangle) } \prod_{i = 0}^{n} \dfrac{\Gamma(\omega[w_i F] + w_i r M \lambda - \langle w_i F \rangle)}{\Gamma(1 -\omega[w_i F] - w_i r M \lambda + \langle w_i F \rangle)} \\
Z_{\text{v}} = & \,z^{-r M \lambda} \sum_{l \geq 0} \dfrac{(rM)^{\vert w \vert l + \sum_{i=0}^n (\omega[w_i F] + [w_i F])} z^{l + F}}{\prod_{i = 0}^{n} (1 -\omega[w_i F] - w_i r M \lambda + \langle w_i F \rangle)_{w_i l + [w_i F] + \omega[w_i F]}}\\
Z_{\text{av}} = & \,\bar{z}^{-r M \lambda} \sum_{k \geq 0} \dfrac{(-rM)^{\vert w \vert k + \sum_{i=0}^n(\omega[w_i F]+ [w_i F])} \bar{z}^{k + F}}{\prod_{i = 0}^{n} (1-\omega[w_i F] - w_i r M \lambda + \langle w_i F \rangle)_{w_i k + [w_i F] +\omega[w_i F]}}\\
\end{split}
\end{equation}
In the formulae we defined $\langle w_i F \rangle$ and $[w_i F]$ as the fractional and integer part of the number $w_i F$, so that $w_i F = [w_i F] +  \langle w_i F \rangle$, while $\vert w \vert = \sum_{i=0}^n w_i $. Moreover, 
\begin{equation}
\omega[w_i F]= \left\{ 
\begin{array}{cc}
0 & \,\,\text{for}\,\, \langle w_i F \rangle = 0\\
1 & \,\,\text{for}\,\, \langle w_i F \rangle \neq 0
\end{array}
\right.
\end{equation}
This is needed in order for the $\mathcal{J}$ function to start with one in the $rM$ expansion. \\

The twisted sectors in \eqref{effe} label the base of the orbifold cohomology space.

Once more, we can also consider complete intersections in $\mathbb{P}^{\textbf{w}}$ of type $(q_0, \ldots, q_m)$. The integrand in \eqref{wpspa} has to be multiplied by
\beq
\prod_{j=0}^m \frac{\Gamma\left(\frac{R_j}{2}-q_j \tau + q_j \frac{m}{2}\right)}{\Gamma\left(1-\frac{R_j}{2} + q_j \tau + q_j \frac{m}{2}\right)}
\eeq
The poles do not change, and
\bea
\begin{split}
Z_{\text{1l}} = & \,(rM)^{- 2 (\vert w \vert - \vert q \vert) r M \lambda - 2 \sum_{i=0}^n (\omega[w_i F] - \langle w_i F \rangle) - 2 \sum_{j=0}^m \langle q_j F \rangle} \\
&\,\prod_{i = 0}^{n} \dfrac{\Gamma(\omega[w_i F]+ w_i r M \lambda - \langle w_i F \rangle)}{\Gamma(1 - \omega[w_i F] - w_i r M \lambda + \langle w_i F \rangle)} 
\prod_{j = 0}^{m} \dfrac{\Gamma(\frac{R_j}{2}-q_j r M \lambda + \langle q_j F \rangle)}{\Gamma(1-\frac{R_j}{2} + q_j r M \lambda - \langle q_j F \rangle)}\\
Z_{\text{v}} = &\, z^{-r M \lambda} \sum_{l \geq 0} (-1)^{\vert q \vert l + \sum_{j=0}^m [q_j F]}(rM)^{(\vert w \vert - \vert q \vert ) l + \sum_{i=0}^n (\omega[w_i F] + [w_i F]) - \sum_{j=0}^m [q_j F]} z^{l + F} \\
&\, \dfrac{\prod_{j = 0}^{m} (\frac{R_j}{2}- q_j r M \lambda + \langle q_j F \rangle)_{q_j l + [q_j F]}}{\prod_{i = 0}^{n} (1 -\omega[w_i F] - w_i r M \lambda + \langle w_i F \rangle)_{w_i l + [w_i F]+ \omega[w_i F]}}\\ 
Z_{\text{av}} = &\, \bar{z}^{-r M \lambda} \sum_{k \geq 0} (-1)^{\vert q \vert k + \sum_{j=0}^m [q_j F]}(-rM)^{(\vert w \vert - \vert q \vert ) k + \sum_{i=0}^n (\omega[w_i F] + [w_i F]) - \sum_{j=0}^m [q_j F]} \bar{z}^{k + F}  \\
& \,\dfrac{\prod_{j = 0}^{m} (\frac{R_j}{2}- q_j r M \lambda + \langle q_j F \rangle)_{q_j k + [q_j F]}}{\prod_{i = 0}^{n} (1 - \omega[w_i F] - w_i r M \lambda + \langle w_i F \rangle)_{w_i k + [w_i F] + \omega[w_i F]}} \label{bunwei}\\
\end{split}
\eea
\vspace{.3cm}

Notice that the non linear sigma model to which the GLSM flows in the IR is well defined only for $\vert w \vert \geq \vert q \vert$, which means for manifolds with $c_1 \geq 0$.

\subsection{Quintic}

We will now consider the most famous compact Calabi-Yau threefold, i.e. the quintic. The corresponding GLSM is a $U(1)$ gauge theory with five chiral fields $\Phi_a$ of charge $+1$, one chiral field $P$ of charge $-5$ and a superpotential of the form $W=PG(\Phi_1,\ldots,\Phi_5)$, where $G$ is a homogeneous polynomial of degree five. We choose the vector R-charges to be $2q$ for the $\Phi$ fields and $(2-5\cdot2q)$ for $P$ such that the superpotential has R-charge $2$. The quintic threefold is realized in the geometric phase corresponding to $\xi>0$. For details of the construction see \cite{Witten:1993yc} and for the relation to the two-sphere partition function \cite{Jockers:2012dk}. Here we want to investigate the connection to the Givental formalism. For a Calabi-Yau manifold the sum of gauge charges is zero, which implies $\xi_{\text{ren}} = \xi$, and $\theta_{\text{ren}} = \theta$ holds because the gauge group is abelian. The spherical partition function is
\begin{equation}
Z \,=\, \sum_{m\in\mathbb{Z}}  \int_{i\mathbb{R}} \dfrac{d \tau}{2 \pi i} z^{-\tau-\frac{m}{2}}\bar{z}^{-\tau+\frac{m}{2}} \left( \dfrac{\Gamma\left(q + \tau - \frac{m}{2}\right)}{\Gamma\left(1-q - \tau - \frac{m}{2}\right)} \right)^5 \dfrac{\Gamma\left(1-5q - 5 \tau +5 \frac{m}{2}\right)}{\Gamma\left(5q + 5 \tau +5 \frac{m}{2}\right)}.
\end{equation}
Since we want to describe the phase $\xi>0$, we have to close the contour in the left half plane. We use the freedom in $q$ to separate the towers of poles coming from $\Phi$'s and from $P$. In the range $0<q<\frac{1}{5}$ the former lie in the left half plane while the latter in the right half plane. So we pick only the poles corresponding to $\Phi$'s given by
\begin{equation}
\tau_k=-q-k+\frac{m}{2},\qquad k\geq \textrm{max}(0,m)
\end{equation}
Then the partition function turns into a sum of residues and we express each residue by the Cauchy contour integral. Finally we arrive at
\begin{equation}
\label{eq:I pairing}
Z=\, (z \bar{z})^{q} \oint_{\mathcal{C}(\delta)} \dfrac{d (r M \lambda)}{2 \pi i} 
Z_{\text{1l}}(\lambda , r M) Z_{\text{v}}(\lambda , r M;z)Z_{\text{av}}(\lambda , r M;\bar{z}),
\end{equation}
where the contour $\mathcal{C}(\delta)$ goes around $\lambda=0$ and
\begin{equation}
\begin{split}
Z_{\text{1l}}(\lambda,r M) &= \dfrac{\Gamma(1-5 r M \lambda)}{\Gamma(5 r M \lambda)}\left(\dfrac{\Gamma(r M \lambda)}{\Gamma(1-r M \lambda)}\right)^{5}\\
Z_{\text{v}}(\lambda,r M;z) &= z^{-r M \lambda}\sum_{l\geqslant 0} (-z)^{l} \dfrac{(1-5 r M \lambda)_{5l}}{[(1-r M \lambda)_l]^5}\\
Z_{\text{av}}(\lambda , r M;\bar{z}) &= \bar{z}^{-r M \lambda}\sum_{k\geqslant 0} (-\bar{z})^{k} \dfrac{(1-5 r M \lambda)_{5k}}{[(1-r M \lambda)_k]^5}\\
\end{split}
\end{equation}
The vortex function $Z_{\text{v}}(\lambda,r M;z)$ reproduces the known Givental $\mathcal{I}$-function
\begin{equation}
\mathcal{I}(H,\hbar ; t) = \sum_{d\geqslant 0} e^{(H/\hbar +d)t} \dfrac{(1+5H/\hbar)_{5d}}{[(1+H/\hbar)_d]^5}
\end{equation}
after identifying
\begin{equation}
H = -\lambda \;\;\;,\;\;\; \hbar = \dfrac{1}{r M} \;\;\;,\;\;\; t = \ln (-z).
\end{equation}
The $\mathcal{I}$-function is valued in cohomology, where $H\in H^2(\mathbb{P}^4)$ is the hyperplane class in the cohomology ring of the embedding space. Because of dimensional reasons we have $H^5=0$ and hence the $\mathcal{I}$-function is a polynomial of order four in $H$
\begin{equation}
\mathcal{I}=I_0+\frac{H}{\hbar}I_1+\left(\frac{H}{\hbar}\right)^2I_2+\left(\frac{H}{\hbar}\right)^3I_3+\left(\frac{H}{\hbar}\right)^4I_4.
\end{equation}
This is naturally encoded in the explicit residue evaluation of \eqref{eq:I pairing}, 
see eq.\eqref{senzanome}.
Now consider the Picard-Fuchs operator $L$. It can be easily shown that $\{I_0,I_1,I_2,I_3\}\in \textrm{Ker}(L)$ while $I_4\notin\textrm{Ker}(L)$. $L$ is an order four operator and so $\mathbf{I}=(I_0,I_1,I_2,I_3)^T$ form a basis of solutions. There exists another basis formed by the periods of the holomorphic $(3,0)$ form of the mirror manifold. In homogeneous coordinates they are given as $\mathbf{\Pi}=(X^0,X^1,\frac{\partial F}{\partial X^0},\frac{\partial F}{\partial X^1})^T$ with $F$ the prepotential. Thus there exists a transition matrix $\mathbf{M}$ relating these two bases
\begin{equation}
\mathbf{I}=\mathbf{M}\cdot\mathbf{\Pi}
\end{equation}
There are now two possible ways to proceed. One would be fixing the transition matrix using mirror construction (i.e. knowing explicitly the periods) and then showing that the pairing given by the contour integral in \eqref{eq:I pairing} after being transformed to the period basis gives the standard formula for the K\"ahler potential in terms of a symplectic pairing
\begin{equation}
\label{eq:Kahler potential}
e^{-K}=i\mathbf{\Pi}^\dagger\cdot\mathbf{\Sigma}\cdot\mathbf{\Pi}
\end{equation}
with $\mathbf{\Sigma}=\begin{pmatrix}
              \mathbf{0}&&\mathbf{1}\\
	      \mathbf{-1}&&\mathbf{0}
             \end{pmatrix}$
being the symplectic form. The other possibility would be to use the fact that the two sphere partition function computes the K\"ahler potential \cite{Jockers:2012dk} and then impose equality between \eqref{eq:I pairing} and \eqref{eq:Kahler potential} to fix the transition matrix. We follow this route in the following. The contour integral in \eqref{eq:I pairing} expresses the K\"ahler potential as a pairing in the $\mathbf{I}$ basis. It is governed by $Z_{\text{1l}}$ which has an expansion
\begin{equation}
Z_{\text{1l}} = \dfrac{5}{(r M \lambda)^4} + \dfrac{400\, \zeta(3)}{r M \lambda} + o(1)
\label{senzanome}
\end{equation}
and so we get after integration (remember that $H/\hbar = - r M \lambda$)
\begin{equation}
\begin{split}
Z&=-2\chi\zeta(3) I_0 \bar{I}_0 - 5 (I_0 \bar{I}_3 + I_1 \bar{I}_2 + I_2 \bar{I}_1 + I_3 \bar{I}_0)\\
&=\mathbf{I}^\dagger\cdot\mathbf{A}\cdot\mathbf{I},
\end{split}
\end{equation}
where
\begin{equation}
\mathbf{A}=\begin{pmatrix}
              -2\chi \zeta(3)&& 0 && 0 && -5\\
	      0 && 0 && -5 && 0\\
	      0 && -5 && 0 && 0\\
	      -5 && 0 && 0 && 0
             \end{pmatrix}
\end{equation}
gives the pairing in the $\mathbf{I}$ basis and $\chi=-200$ is the Euler characteristic of the quintic threefold.
From the two expressions for the K\"ahler potential we easily find the transition matrix as
\begin{equation}
\mathbf{M}=\begin{pmatrix}
              1 && 0 && 0 && 0\\
	      0 && 1 && 0 && 0\\
	      0 && 0 && 0 && -\frac{i}{5}\\
	      -\frac{\chi}{5}\zeta(3) && 0 && -\frac{i}{5} && 0
             \end{pmatrix}.
\end{equation}
Finally, we know that the mirror map is given by 
\begin{equation}
t = \dfrac{I_1}{2\pi i I_0} \;\;\;,\;\;\; \bar{t} = -\dfrac{\bar{I}_1}{2\pi i \bar{I}_0}
\end{equation}
so after dividing $Z$ by $ (2\pi i)^2 I_0 \bar{I}_0$ for the change of coordinates and by a further $2\pi$ for the normalization of the $\zeta(3)$ term, we obtain the K\"ahler potential in terms of $t$, $\bar{t}$, in a form in which the symplectic product is evident.

\subsection{Local Calabi--Yau: $\mathcal{O}(p)\oplus \mathcal{O}(-2-p) \rightarrow \mathbb{P}^{1}$}

Let us now study the family of spaces $X_{p} = \mathcal{O}(p)\oplus \mathcal{O}(-2-p) \rightarrow \mathbb{P}^{1}$ with diagonal equivariant action on the fiber. We will find exact agreement with the $\mathcal{I}$ functions computed in \cite{2006math......3728F}, and we will show how the quantum corrected K\"ahler potential for the K\"ahler moduli space can be computed when equivariant parameters are turned on.\\
Here we will restrict only to the phase $\xi > 0$, which is the one related to $X_{p}$. The case $\xi < 0$ describes the orbifold phase of the model; this will be studied in the following sections.

\subsubsection{Case $p = -1$}

First of all, we have to write down the partition function; this is given by
\begin{equation}
Z_{-1} = \sum_{m\in\mathbb{Z}} e^{-i m \theta} \int \dfrac{d \tau}{2 \pi i} e^{4\pi \xi \tau} \left( \dfrac{\Gamma\left(\tau - \frac{m}{2}\right)}{\Gamma\left(1 - \tau - \frac{m}{2}\right)} \right)^2  \left( \dfrac{\Gamma\left( - \tau - i r M a + \frac{m}{2}\right)}{\Gamma\left(1 + \tau + i r M a + \frac{m}{2}\right)} \right)^2 \label{-1}
\end{equation}
The poles are located at
\begin{equation}
\tau = -k + \dfrac{m}{2} + r M \lambda \label{pole-1}
\end{equation}
so we can rewrite \eqref{-1} as
\begin{equation}
Z_{-1} = \oint \dfrac{d (r M \lambda)}{2 \pi i} Z_{\text{1l}} Z_{\text{v}} Z_{\text{av}}  
\end{equation}
where
\begin{equation}
\begin{split}
Z_{\text{1l}} =\,& \left(\dfrac{\Gamma(r M \lambda)}{\Gamma(1-r M \lambda)} \dfrac{\Gamma(-r M \lambda - i r M a)}{\Gamma(1+r M \lambda + i r M a )}\right)^{2} \\
Z_{\text{v}} =\,& z^{-r M \lambda} \sum_{l\geqslant 0} z^{l} \dfrac{(- r M \lambda - i r M a)_{l}^2}{(1-r M \lambda)_l^2} \\
Z_{\text{av}} =\,& \bar{z}^{-r M \lambda} \sum_{k\geqslant 0} \bar{z}^{k} \dfrac{(- r M \lambda - i r M a)_{k}^2}{(1-r M \lambda)_k^2} \\
\end{split}
\end{equation}
Notice that our vortex partition function coincides with the Givental function given in \cite{2006math......3728F}
\begin{equation}
\mathcal{I}_{-1}^T(q) = e^{\frac{H}{\hbar} \ln q } \sum_{d\geqslant 0} \dfrac{(1-H/\hbar + \tilde{\lambda} /\hbar -d)_d^2}{(1+H/\hbar)_d^2} q^d
\end{equation}
after the usual identifications
\begin{equation}
H = -\lambda \;\;\;,\;\;\; \hbar = \frac{1}{r M} \;\;\;,\;\;\; \tilde{\lambda} = i a \;\;\;,\;\;\; q = z \label{map1}
\end{equation}
Now, expanding $\mathcal{I}_{-1}^T$ in $r M = 1/\hbar$ we find
\begin{equation}
\mathcal{I}_{-1}^T= 1 - r M \lambda \log z + o((r M)^2)
\end{equation}
which means the mirror map is trivial and the equivariant mirror map absent, i.e. $\mathcal{I}_{-1}^T = \mathcal{J}_{-1}^T$. What remains to be specified is the normalization of the 1-loop factor. 
As explained in \cite{Bonelli:2013rja}, this normalization is fixed by requiring the cancellation of 
the Euler-Mascheroni constants appearing in the Weierstrass form of the $\Gamma$-function, 
reproduces the classical intersection numbers and starts from 1 in the $rM$ expansion; 
in our case, the factor 
\begin{equation}
(z \bar{z})^{-i r M a /2} \left(\dfrac{\Gamma(1+ i r M a)}{\Gamma(1 - i r M a)}\right)^{2} \label{1lo}
\end{equation}
does the job. We can now integrate in $r M \lambda$ and expand in $r M$, obtaining (for $r M a = i q$)
\begin{equation}
\begin{split}
Z_{-1} =&\, \dfrac{2}{q^3} -\dfrac{1}{4 q}\ln^{2} (z \bar{z}) + \Big[ -\dfrac{1}{12} \ln^{3} (z \bar{z}) - \ln (z \bar{z})(\text{Li}_2(z) + \text{Li}_2(\bar{z}))\\ 
& \,+ 2 (\text{Li}_3(z) + \text{Li}_3(\bar{z})) + 4 \zeta(3) \Big] + o(rM) \\
\end{split}
\end{equation}
The terms inside the square brackets reproduce the K\"ahler potential we are interested in, once we multiply everything by $\frac{1}{2\pi (2\pi i)^2}$ and define
\begin{equation}
t = \dfrac{1}{2\pi i}\ln z \;\;\;,\;\;\; \bar{t} = -\dfrac{1}{2\pi i}\ln \bar{z}.
\end{equation} 

\subsubsection{Case $p = 0$}

In this case case, the spherical partition function is
\begin{equation}
Z_{0} = \sum_{m\in\mathbb{Z}} e^{-i m \theta} \int \dfrac{d \tau}{2 \pi i} e^{4\pi \xi \tau} \left( \dfrac{\Gamma\left(\tau - \frac{m}{2}\right)}{\Gamma\left(1 - \tau - \frac{m}{2}\right)} \right)^2   \dfrac{\Gamma\left(-i r M a \right)}{\Gamma\left(1 + i r M a \right)}\dfrac{\Gamma\left( - 2 \tau -i r M a + 2\frac{m}{2}\right)}{\Gamma\left(1 + 2 \tau + i r M a + 2 \frac{m}{2}\right)}
\end{equation}
The poles are as in \eqref{pole-1}, and usual manipulations result in
\begin{equation}
\begin{split}
Z_{\text{1l}} =\,& \left(\dfrac{\Gamma(r M \lambda)}{\Gamma(1-r M \lambda)}\right)^{2}\dfrac{\Gamma\left(-i r M a \right)}{\Gamma\left(1 + i r M a \right)}\dfrac{\Gamma(-2 r M \lambda - i r M a)}{\Gamma(1+2 r M \lambda + i r M a)} \\
Z_{\text{v}} =\,& z^{-r M \lambda} \sum_{l\geqslant 0} z^{l} \dfrac{(- 2 r M \lambda - i r M a )_{2 l}}{(1-r M \lambda)_l^2} \\
Z_{\text{av}} =\,& \bar{z}^{-r M \lambda} \sum_{k\geqslant 0} \bar{z}^{k } \dfrac{(- 2 r M \lambda - i r M a )_{2 k}}{(1-r M \lambda)_k^2} \\
\end{split}
\end{equation}
Again, we recover the Givental function
\begin{equation}
\mathcal{I}_{0}^T(q) = e^{\frac{H}{\hbar} \ln q } \sum_{d\geqslant 0} \dfrac{(1-2 H/\hbar + \tilde{\lambda} /\hbar - 2 d)_{2 d}}{(1+H/\hbar)_d^2} q^d
\end{equation}
of \cite{2006math......3728F} under the map \eqref{map1}; its expansion in $r M$ 
\begin{equation}
\mathcal{I}_{0}^T = 1 - r M \lambda \left[ \log z + 2 \sum_{k=1}^{\infty} z^k \frac{\Gamma(2k)}{(k!)^2} \right] -i r M a \sum_{k=1}^{\infty} z^k \frac{\Gamma(2k)}{(k!)^2}  + o((r M)^2)
\end{equation}
implies that the mirror map is (modulo $(2 \pi i)^{-1}$)
\begin{equation}
t = \log z + 2 \sum_{k=1}^{\infty} z^k \frac{\Gamma(2k)}{(k!)^2}
\end{equation}
and the equivariant mirror map is 
\begin{equation}
\tilde{t} = \frac{1}{2}(t - \log z) = \sum_{k=1}^{\infty} z^k \frac{\Gamma(2k)}{(k!)^2}
\end{equation}
The $\mathcal{J}$ function can be recovered by inverting the equivariant mirror map and changing coordinates accordingly, that is 
\begin{equation}
\mathcal{J}_{0}^T(t) = e^{i r M a \tilde{t}(z)} \mathcal{I}_{0}^T(z) = e^{i r M a \tilde{t}(z)} Z_{\text{v}}(z)
\end{equation}
A similar job has to be done for $Z_{\text{av}}$. The normalization for the 1-loop factor is the same as \eqref{1lo} but in $t$ coordinates, which means 
\begin{equation}
(t \bar{t})^{-i r M a /2} \left(\dfrac{\Gamma(1+ i r M a)}{\Gamma(1 - i r M a)}\right)^{2};
\end{equation}
Finally, integrating in $r M \lambda$ and expanding in $r M$ we find
\begin{equation}
\begin{split}
Z_{0} =&\, \dfrac{2}{q^3} -\dfrac{1}{4 q}(t + \bar{t})^{2}  + \Big[  -\dfrac{1}{12} (t + \bar{t})^{3} - (t + \bar{t})(\text{Li}_2(e^t) + \text{Li}_2(e^{\bar{t}})) \\ 
& \, + 2 (\text{Li}_3(e^t) + \text{Li}_3(e^{\bar{t}})) + 4 \zeta(3) \Big] + o(rM) \\
\end{split}
\end{equation}
As it was shown in \cite{2006math......3728F}, this proves that the two Givental functions $\mathcal{J}_{-1}^T$ and $\mathcal{J}_{0}^T$ are the same, as well as the K\"ahler potentials; the $\mathcal{I}$ functions look different simply because of the choice of coordinates on the moduli space.

\subsubsection{Case $p \geq 1 $}

In the general $p \geq 1$ case, we have 
\begin{equation}
\begin{split}
Z_p =&\, \sum_{m\in\mathbb{Z}} e^{-i m \theta} \int \dfrac{d \tau}{2 \pi i} e^{4\pi \xi \tau} \left( \dfrac{\Gamma\left(\tau - \frac{m}{2}\right)}{\Gamma\left(1 - \tau - \frac{m}{2}\right)} \right)^2 \\
& \, \dfrac{\Gamma\left( - (p+2) \tau -i r M a + (p+2)\frac{m}{2}\right)}{\Gamma\left(1 + (p+2) \tau + i r M a + (p+2) \frac{m}{2}\right)} \dfrac{\Gamma\left( p \tau -i r M a -p\frac{m}{2}\right)}{\Gamma\left(1 -p \tau + i r M a -p \frac{m}{2}\right)} \\ \label{zp}
\end{split}
\end{equation}
There are two classes of poles, given by
\begin{eqnarray}
\tau &=& -k + \dfrac{m}{2} + r M \lambda   \label{set1} \\
\tau &=& -k + \dfrac{m}{2} + r M \lambda - F + i r M \dfrac{a}{p} \label{set2}
\end{eqnarray}
where $F = \{0, \frac{1}{p}, \ldots , \frac{p-1}{p}\}$ and the integration is around $r M \lambda = 0$. 
This can be understood from the fact that actually the GLSM \eqref{zp} describes the canonical bundle over the weighted projective space $\mathbb{P}_{(1,1,p)}$, which has two chambers.
The regular one, associated to the poles \eqref{set1}, corresponds to the local $\mathcal{O}(p)\oplus \mathcal{O}(-2-p) \rightarrow \mathbb{P}^{1}$ geometry:
\begin{equation}
Z_p^{(0)} = \oint \dfrac{d (r M \lambda)}{2 \pi i} Z_{\text{1l}}^{(0)} Z_{\text{v}}^{(0)} Z_{\text{av}}^{(0)}  
\end{equation}
with
\begin{equation}
\begin{split}
Z_{\text{1l}}^{(0)} =\,& \left(\dfrac{\Gamma(r M \lambda)}{\Gamma(1-r M \lambda)}\right)^{2}\dfrac{\Gamma(-(p+2) r M \lambda - i r M a )}{\Gamma(1+ (p+2)r M \lambda  + i r M a)}\dfrac{\Gamma(p\, r M \lambda - i r M a)}{\Gamma(1 - p\, r M \lambda + i r M a)} \\
Z_{\text{v}}^{(0)} =\,& z^{-r M \lambda} \sum_{l\geqslant 0} (-1)^{(p+2)l} z^{l} \dfrac{(-(p+2) r M \lambda - i r M a)_{(p+2) l}}{(1-r M \lambda)_l^2(1 - p\, r M \lambda + i r M a)_{p l}} \\
Z_{\text{av}}^{(0)} =\,& \bar{z}^{-r M \lambda} \sum_{k\geqslant 0} (-1)^{(p+2)k} \bar{z}^{k} \dfrac{(-(p+2) r M \lambda - i r M a)_{(p+2) k}}{(1-r M \lambda)_k^2(1 - p\, r M \lambda + i r M a)_{p k}} \\
\end{split}
\end{equation}
The second chamber, associated to \eqref{set2}, is an orbifold one:
\begin{equation}
Z_p^{(F)} = \sum_{\delta = 0}^{p-1} \oint \dfrac{d (r M \lambda)}{2 \pi i} Z_{\text{1l},\delta}^{(F)} Z_{\text{v},\delta}^{(F)} Z_{\text{av},\delta}^{(F)} \label{F}
\end{equation}
where $F = \frac{\delta}{p}$. The explicit expression for $Z^{(F)}$ in the above formula can be recovered from \eqref{bunwei}, adding the twisted masses in the appropriate places. Notice that \eqref{F} can be easily integrated, since there are just simple poles.

\subsection{Orbifold Gromov-Witten invariants}
In this section we want to show how the analytic structure of the partition function encodes all the classical phases of the abelian GLSM. These are given by the secondary fan, which in our conventions is generated by the columns of the charge matrix $Q$. In terms of the partition function these phases are governed by the choice of integration contours, namely by the structure of poles we are picking up. 
The contour can be closed either in the left half plane (for $\xi>0$) or in the right half plane ($\xi<0$)\footnote{This is only true for Calabi-Yau manifolds; for $c_1 > 0$, i.e. $\sum_i Q_i >0$, the contour is fixed.}. 
The transition between different phases occurs when some of the integration contours are flipped 
and the corresponding variable is integrated. 
To summarize, a single partition function contains the ${\cal I}$-functions of geometries corresponding 
to all the different phases of the GLSM. These geometries are related by minimally resolving the 
singularities by blow-up until the complete smoothing of the space takes place (when this is possible). 
Our procedure consists in considering the GLSM corresponding to the complete resolution and
its partition function.
Then by flipping contours and doing partial integrations one discovers all other, more singular geometries. 
In the following we illustrate these ideas on a couple of examples.

\subsubsection{$K_{\mathbb{P}^{n-1}}$ vs. $\mathbb{C}^n/\mathbb{Z}_n$}

Let us consider a $U(1)$ gauge theory with $n$ chiral fields of charge $+1$ and one chiral field of charge $-n$. The secondary fan is generated by two vectors $\{1,-n\}$ and so has two chambers corresponding to two different phases.  For $\xi > 0$ it describes a smooth geometry $K_{\mathbb{P}^{n-1}}$, that is the total space of the canonical bundle over the complex projective space ${\mathbb{P}^{n-1}}$, while for $\xi<0$ the orbifold 
$\mathbb{C}^n/\mathbb{Z}_n$. The case $n=3$ will reproduce the results of 
\cite{Aganagic:2006wq,2007math......2234C,2008arXiv0804.2592C}. The partition function reads
\begin{equation}
Z = \sum_m \int_{i\mathbb{R}} \dfrac{d \tau}{2 \pi i} e^{4 \pi \xi \tau - i \theta m} \left( \dfrac{\Gamma(\tau - \frac{m}{2})}{\Gamma(1 - \tau - \frac{m}{2})} \right)^n \dfrac{\Gamma(-n \tau +n \frac{m}{2} + i r M a)}{\Gamma(1 +n \tau +n \frac{m}{2} - i r M a)}
\end{equation}
Closing the contour in the left half plane (i.e. for $\xi > 0$) we take poles at
\begin{equation}
\tau = -k + \frac{m}{2} + r M \lambda
\end{equation}
and obtain
\begin{equation}
\begin{split}
Z =& \, \oint \dfrac{d (r M \lambda)}{2 \pi i}  \left( \dfrac{\Gamma(r M \lambda)}{\Gamma(1 - r M \lambda)} \right)^n \dfrac{\Gamma(-n r M \lambda + i r M a)}{\Gamma(1 +n r M \lambda - i r M a)} \\
&\, \sum_{l \geq 0} z^{-r M \lambda} (-1)^{nl}z^{n l} \dfrac{(-nr M \lambda + i r M a)_{nl}}{(1-r M \lambda)^n_l} \\
&\, \sum_{k \geq 0} \bar{z}^{-r M \lambda} (-1)^{nk}\bar{z}^{n k} \dfrac{(-nr M \lambda + i r M a)_{nk}}{(1-r M \lambda)^n_k} \\
\end{split}
\end{equation}
We thus find exactly the Givental function for $K_{\mathbb{P}^{n-1}}$. To switch to the singular geometry we flip the contour and do the integration. Closing in the right half plane ($\xi < 0$) we consider
\begin{equation}
\tau = k + \dfrac{\delta}{n} + \frac{m}{2} + \dfrac{1}{n}i r M a
\end{equation}
with $\delta = 0,1,2, \ldots, n-1$.  After integrating over $\tau$, we obtain
\begin{equation}
\begin{split}
Z &=\, \dfrac{1}{n} \sum_{\delta = 0}^{n-1} \left( \dfrac{\Gamma(\frac{\delta}{n} + \frac{1}{n}i r M a )}{\Gamma(1 - \frac{\delta}{n} - \frac{1}{n}i r M a )} \right)^n \dfrac{1}{(rM)^{2 \delta}} \\
&\, \sum_{k \geq 0} (-1)^{n k} (\bar{z}^{-1/n})^{nk + \delta + i r M a} (rM)^{\delta} \dfrac{(\frac{\delta}{n} + \frac{1}{n}i r M a )^n_k}{(nk + \delta)!} \\
&\, \sum_{l \geq 0} (-1)^{n l} (z^{-1/n})^{nl + \delta + i r M a} (-rM)^{\delta} \dfrac{(\frac{\delta}{n} + \frac{1}{n}i r M a )^n_l}{(nl + \delta)!} \\
\end{split}
\end{equation}
as expected from \eqref{bunwei}. Notice that when the contour is closed in the right half plane, vortex and antivortex contributions are exchanged. 
We can compare the $n=3$ case corresponding to $\mathbb{C}^3/\mathbb{Z}_3$  with \cite{2008arXiv0804.2592C}, 
given by
\begin{equation}
\mathcal{I} = x^{-\lambda / z} \sum_{\substack{d \in \mathbb{N} \\ d\geq 0}} \dfrac{x^d}{d! z^d} \prod_{\substack{ 0 \leq b < \frac{d}{3} \\ \langle b \rangle = \langle \frac{d}{3} \rangle }} \left( \frac{\lambda}{3} - b z \right)^3 \mathbf{1}_{\langle \frac{d}{3} \rangle} 
\end{equation}
which in a more familiar notation becomes
\begin{equation}
\mathcal{I} = x^{-\lambda / z} \sum_{\substack{d \in \mathbb{N} \\ d\geq 0}} \dfrac{x^d}{d!} \dfrac{1}{z^{3 \langle \frac{d}{3} \rangle }} (-1)^{3 [\frac{d}{3}]} \left( \langle \frac{d}{3} \rangle -\frac{\lambda}{3 z} \right)^3_{[\frac{d}{3}]} \mathbf{1}_{\langle \frac{d}{3} \rangle} 
\end{equation}
The necessary identifications are straightforward.

\subsubsection{The quantum cohomology of 
$\mathbb{C}^3/\mathbb{Z}_{p+2}$
and its crepant resolution}

We now consider the orbifold space $\mathbb{C}^3/\mathbb{Z}_{p+2}$ with weights $(1,1,p)$ and $p>1$.
Its full crepant resolution is provided by a resolved transversal $A_{p+1}$ singularity 
(namely a local Calabi-Yau threefold obtained by fibering the resolved $A_{p+1}$ singularity over 
a $\mathbb{P}^1$ base space). 
The corresponding GLSM contains 
$p + 2$ abelian gauge groups and $p+5$ chiral multiplets, with the following charge assignment:
\begin{equation}
\left( \begin{array}{cccccccccccc}
  0\;  & \; 1 \; & \;1\; & \;-1\; & \;-1\; & \;0\; & \;\ldots \;& \;0\; & \;0\; & \;0\; & \;\ldots \;& \;0\\
-j-1 \;& \; j \; &\; 0\; & \;0 \; & \; 0\; &\; 0\; &\; \ldots \;& \;0\; &\; \stackrel{(5+j){\rm th}}{1} \;& \;0\; &\; \ldots \;&\; 0\\ 
 -p-2 \; &\; p+1\; &\; 1\; & \;0 \; & \; 0\; & \;0\; &\; \ldots \;& \;0\; &\; 0\; &\; 0\; &\; \ldots \; &\; 0
\end{array} \right) \label{cm}
\end{equation}
where $1 \leq j \leq p$. 
In the following we focus on the particular chambers corresponding to  the partial resolutions 
$K_{\mathbb{F}_p}$ and $K_{\mathbb{P}^2(1,1,p)}$. 
Let us start by discussing the local $\mathbb{F}_p$ chamber: 
this can be seen by replacing the last row in \eqref{cm} with the linear combination
\begin{equation}
(\text{last row}) \;\;\longrightarrow \;\; (\text{last row}) - p\, (\text{second row}) - (\text{first row} ) 
\end{equation} 
which corresponds to
\begin{equation}
\left( \begin{array}{ccccccc} -p-2\; & \;p+1\; & \;1 \;& \;0\; & \;0\; & \; 0 \; & \;\ldots \end{array} \right)
\longrightarrow
\left( \begin{array}{ccccccc} p-2\; & \;0\; & \;0 \;& \;1\; & \;1\; & \; -p \; & \;\ldots \end{array} \right)
\end{equation}
The charge matrix \eqref{cm} now reads ($2 \leq n \leq p$)
\begin{equation}
\left( \begin{array}{cccccccccccc}
0\;  & \; 1 \; & \;1\; & \;-1\; & \;-1\; & \;0\; & \;\ldots \;& \;0\; & \;0\; & \;0\; & \;\ldots \;& \;0\\
-2\; & \;1\; & \;0 \;& \;0\; & \;0\; & \; 1 \; &\; \ldots \;& \;0\; &\; 0\; &\; 0\; &\; \ldots \; &\; 0\\
-n-1 \;& \; n \; &\; 0\; & \;0 \; & \; 0\; &\; 0\; &\; \ldots \;& \;0\; &\; \stackrel{(5+n){\rm th}}{1} \;& \;0\; &\; \ldots \;&\; 0\\ 
 p-2\; & \;0\; & \;0 \;& \;1\; & \;1\; & \; -p \; &\; \ldots \;& \;0\; &\; 0\; &\; 0\; &\; \ldots \; &\; 0
\end{array} \right) \label{new}
\end{equation}
and, in a particular sector (i.e. for a particular choice of poles), after turning to infinity $p$ Fayet-Iliopoulos parameters, we remain with the second and the last row:
\begin{equation}
Q=\left( \begin{array}{ccccc}
-2 \;& \; 1 \; & \;0 \; & \; 0\; &\; 1 \\ 
 p-2\; & \;0\; & \;1\; & \;1\; & \; -p 
\end{array} \right) \label{this}
\end{equation}
which is the charge matrix of $K_{\mathbb{F}_p}$.

Let us see how this happens in detail; since it is easier for our purposes, we will consider the charge 
matrix \eqref{new}. For generic $p$, the partition function with the addition of a twisted mass for the field corresponding to the first column of \eqref{new} is given by 
\begin{equation}
\begin{split}
Z =&\, \sum_{m_0, \ldots , m_{p+1}} \oint \left[ \prod_{i = 0}^{p+1} \dfrac{d \tau_i}{2\pi i} z_i^{-\tau_i -\frac{m_i}{2}}\bar{z}_i^{-\tau_i +\frac{m_i}{2}} \right] \left[ \prod_{j = 0}^{p} \dfrac{\Gamma(\tau_j - \frac{m_j}{2})}{\Gamma(1 - \tau_j - \frac{m_j}{2})}  \right] 
\dfrac{\Gamma(\tau_1 - p\tau_{p+1} - \frac{m_1}{2} + p \frac{m_{p+1}}{2})}{\Gamma(1 - \tau_1 + p \tau_{p+1} - \frac{m_1}{2} + p \frac{m_{p+1}}{2})}   \\
&\, \left(\dfrac{\Gamma(-\tau_0 + \tau_{p+1} + \frac{m_0}{2} - \frac{m_{p+1}}{2} )}{\Gamma(1 +\tau_0 - \tau_{p+1} + \frac{m_0}{2} - \frac{m_{p+1}}{2})} \right)^2 
\dfrac{\Gamma(\tau_0 + \sum_{j=1}^{p} j \tau_j - \frac{m_0}{2} - \sum_{j=1}^{p} j \frac{m_j}{2})}{\Gamma(1 -\tau_0 - \sum_{j=1}^{p} j \tau_j - \frac{m_0}{2} - \sum_{j=1}^{p} j \frac{m_j}{2})} \\
&\, \dfrac{\Gamma( -\sum_{j=1}^{p} (j+1) \tau_j  +(p-2) \tau_{p+1} + \sum_{j=1}^{p} (j+1) \frac{m_j}{2} - (p-2) \frac{m_{p+1}}{2} +i r M a)}{\Gamma(1 + \sum_{j=1}^{p} (j+1) \tau_j -(p-2)\tau_{p+1} + \sum_{j=1}^{p} (j+1) \frac{m_j}{2} - (p-2) \frac{m_{p+1}}{2}-i r M a)} \\
\end{split}
\end{equation}
Now, choosing the sector
\begin{equation}
\begin{split}
& \tau_0 = - k_0 + \frac{m_0}{2} \\
& \tau_n = -k_n + \dfrac{m_n}{2} \,\,\,\,\,\,,\,\,\,\,\,\, 2 \leq n \leq p \\
\end{split}
\end{equation}
and integrating over these variables we arrive at
\begin{equation}
\begin{split}
Z =&\, \sum_{k_0, k_{n} \geq 0 } \sum_{l_0, l_{n} \geq 0 } \dfrac{z_0^{l_0}}{l_0!} \dfrac{(-1)^{k_0}\bar{z}_i^{k_0}}{k_0!} \prod_{n=2}^{p} \dfrac{z_i^{l_i}}{l_i!} \dfrac{(-1)^{k_i}\bar{z}_i^{k_i}}{k_i!} \sum_{m_1, m_{p+1}}\\
&\, \oint \dfrac{d \tau_{1}}{2\pi i}\dfrac{d \tau_{p+1}}{2\pi i}e^{4 \pi \xi_{1} \tau_{1} -i \theta_{1} m_{1}} e^{4 \pi \xi_{p+1} \tau_{p+1} -i \theta_{p+1} m_{p+1}}  
\dfrac{\Gamma(\tau_1 - p\tau_{p+1} - \frac{m_1}{2} + p \frac{m_{p+1}}{2})}{\Gamma(1 - \tau_1 - \frac{m_1}{2} + p \tau_{p+1} + p \frac{m_{p+1}}{2})}   \\ 
&\, \left(\dfrac{\Gamma(k_0 + \tau_{p+1} - \frac{m_{p+1}}{2} )}{\Gamma(1 -l_0 - \tau_{p+1} - \frac{m_{p+1}}{2})} \right)^2 
\dfrac{\Gamma(-k_0 + \tau_1 - \sum_{n=2}^{p} n k_n - \frac{m_1}{2} )}{\Gamma(1 +l_0 - \tau_1 + \sum_{n=2}^{p} n l_n  - \frac{m_1}{2})} \\
&\, \dfrac{\Gamma( -2\tau_1 + \sum_{n=2}^{p} (n+1) k_n  + (p-2) \tau_{p+1} + 2 \frac{m_1}{2} - (p-2) \frac{m_{p+1}}{2} +i r M a)}{\Gamma(1 + 2 \tau_2 - \sum_{n=2}^{p} (n+1) l_n -(p-2)\tau_{p+1} + 2 \frac{m_2}{2} - (p-2) \frac{m_{p+1}}{2}-i r M a)} \\
\end{split}
\end{equation}
which defines a linear sigma model with charges \eqref{this} for $k_0 = k_{n} = 0$, $l_0 = l_{n} = 0$ 
(i.e. when $\xi_0 = \xi_n = \infty$). 

The secondary fan of this model has four chambers, but here we concentrate only on three of them, describing $K_{\mathbb{F}_p}$, $K_{\mathbb{P}^2(1,1,p)}$ and $\mathbb{C}^3/\mathbb{Z}_{p+2}$ respectively. Its partition function is given by 
\begin{equation}
\begin{split}
Z =&\, \sum_{m_1, m_{p+1}} \int \dfrac{d \tau_1}{2 \pi i} \dfrac{d \tau_{p+1}}{2 \pi i}  
e^{4\pi \xi_1 \tau_1 - i \theta_1 m_1}  e^{4\pi \xi_{p+1} \tau_{p+1} - i \theta_{p+1} m_{p+1}} 
\left( \dfrac{\Gamma(\tau_{p+1} - \frac{m_{p+1}}{2})}{\Gamma(1 - \tau_{p+1} - \frac{m_{p+1}}{2})} \right)^2 
\dfrac{\Gamma(\tau_1 - \frac{m_1}{2})}{\Gamma(1 - \tau_1 - \frac{m_1}{2})} \\
& \, \dfrac{\Gamma(-p \tau_{p+1} + \tau_1 +p \frac{m_{p+1}}{2} - \frac{m_1}{2})}{\Gamma(1 +p \tau_{p+1} -\tau_1 + p\frac{m_{p+1}}{2} - \frac{m_1}{2})}
\dfrac{\Gamma((p-2) \tau_{p+1} -2 \tau_1 -(p-2) \frac{m_{p+1}}{2} + 2\frac{m_1}{2} + i r M a)}{\Gamma(1 -(p-2) \tau_{p+1} +2\tau_1 - (p-2)\frac{m_{p+1}}{2} +2 \frac{m_1}{2} - i r M a)}
\end{split}
\end{equation} 
If we consider the set of poles 
\begin{eqnarray}
\tau_{p+1} &=& -k_{p+1} + \dfrac{m_{p+1}}{2} + r M \lambda_{p+1} \nonumber\\
\tau_1 &=& -k_1 + \dfrac{m_1}{2} + r M \lambda_1
\end{eqnarray}
we are describing the canonical bundle over $\mathbb{F}_p$:
\begin{equation}
\begin{split}
& Z_{K_{\mathbb{F}_p}} =\, \oint \dfrac{d (r M \lambda_1)}{2 \pi i} \dfrac{d (r M \lambda_{p+1})}{2 \pi i} 
\left( \dfrac{\Gamma(r M \lambda_{p+1})}{\Gamma(1 - r M \lambda_{p+1})} \right)^2 
\dfrac{\Gamma(r M \lambda_1)}{\Gamma(1 - r M \lambda_1)}\\
& \dfrac{\Gamma(-p r M \lambda_{p+1} + r M \lambda_1 )}{\Gamma(1 +p r M \lambda_{p+1} -r M \lambda_1 )}
\dfrac{\Gamma((p-2) r M \lambda_{p+1} -2 r M \lambda_1 + i r M a)}{\Gamma(1 -(p-2) r M \lambda_{p+1} + 2 r M \lambda_1 - i r M a)}\\ 
&\sum_{l_1, l_{p+1}} (-1)^{(p-2)l_{p+1}} z_{p+1}^{l_{p+1} - r M \lambda_{p+1}}z_1^{l_1 - r M \lambda_1} \dfrac{((p-2) r M \lambda_{p+1} -2 r M \lambda_1 + i r M a)_{2l_1 - (p-2)l_{p+1}}}{(1 - r M \lambda_{p+1})^2_{l_{p+1}} (1 - r M \lambda_1)_{l_1}(1 +p r M \lambda_{p+1} -r M \lambda_1 )_{l_1 - p l_{p+1}}}\\
&\sum_{k_1, k_{p+1}} (-1)^{(p-2)k_{p+1}} \bar{z}_{p+1}^{k_{p+1} - r M \lambda_{p+1}}\bar{z}_1^{k_1 - r M \lambda_1} \dfrac{((p-2) r M \lambda_{p+1} -2 r M \lambda_1 + i r M a)_{2k_1 - (p-2)k_{p+1}}}{(1 - r M \lambda_{p+1})^2_{k_{p+1}} (1 - r M \lambda_1)_{k_1}(1 +p r M \lambda_{p+1} -r M \lambda_1 )_{k_1 - p k_{p+1}}} 
\end{split}
\end{equation} 
On the other hand, taking poles for
\begin{equation}
\tau_1 = p\tau_{p+1} -p\dfrac{m_{p+1}}{2} + \dfrac{m_1}{2} - k_1 \label{pole}
\end{equation}
and integrating over $\tau_1$ we obtain the canonical bundle over $\mathbb{P}^2_{(1,1,p)}$:
\begin{equation}
\begin{split}
\label{eq:Z_KP(1,1,p)}
Z_{K_{\mathbb{P}^2_{(1,1,p)}}} =& \, \sum_{k_1, l_1 \geq 0}   
\dfrac{z_1^{l_1}}{l_1!} \dfrac{(-1)^{k_1}\bar{z}_1^{k_1}}{k_1!} \\
&\, \sum_{m_{p+1}} \int \dfrac{d \tau_{p+1}}{2 \pi i}  
e^{4\pi (\xi_{p+1} + p \xi_1) \tau_{p+1} - i (\theta_{p+1} + p \theta_1) m_{p+1}}\left( \dfrac{\Gamma(\tau_{p+1} - \frac{m_{p+1}}{2})}{\Gamma(1 - \tau_{p+1} - \frac{m_{p+1}}{2})} \right)^2 \\
&\,\dfrac{\Gamma(p \tau_{p+1} - p \frac{m_{p+1}}{2} - k_1)}{\Gamma(1 - p \tau_{p+1} - p \frac{m_{p+1}}{2} + l_1)}  \dfrac{\Gamma(-(p+2) \tau_{p+1} + (p+2) \frac{m_{p+1}}{2}+ i r M a + 2 k_1 )}{\Gamma(1 + (p+2) \tau_{p+1} + (p+2) \frac{m_{p+1}}{2} - i r M a - 2 l_1)}
\end{split}
\end{equation}
with $l_1 = k_1 - m_1 + p m_{p+1}$ and $z_1 = e^{-2 \pi \xi_1 + i \theta_1}$. In fact, in the limit $\xi_1 \rightarrow \infty$ with $\xi_{p+1} + p \xi_1$ finite, only the $k_1 = l_1 = 0$ sector contributes, leaving the linear sigma model of $K_{\mathbb{CP}^2_{(1,1,p)}}$ for $\xi_{p+1} + p \xi_1 > 0$.\\
From the point of view of the charge matrix, the choice \eqref{pole} corresponds to take linear combinations of the rows, in particular 
\begin{equation}
\left( \begin{array}{ccccc} p-2\; & \;0\; & \;1 \;& \;1\; & \;-p \end{array} \right)
\longrightarrow
\left( \begin{array}{ccccc} p-2\; & \;0\; & \;1 \;& \;1\; & \;-p \end{array} \right) + p \left( \begin{array}{ccccc} -2\; & \;1\; & \;0 \;& \;0 \;& \;1 \end{array} \right) 
\end{equation}
which implies $\xi_{p+1} \rightarrow \xi_{p+1} + p \xi_1$, $\theta_{p+1} \rightarrow \theta_{p+1} + p \theta_1$ and
\begin{equation}
\left( \begin{array}{ccccc} -2\; & \;1 \;& \;0 \;& \;0\; &\; 1\; \\ 
p-2\; & \;0\; & \;1 \;& \;1\; & \;-p \end{array} \right)
\longrightarrow
\left( \begin{array}{ccccc} -2\; & \;1\; &\; 0\; & \;0 \;& \;1 
\\ -p-2\; & \;p\; & \;1 \;& \;1\; & \;0 \end{array} \right)
\end{equation}
while the process of integrating in $ \tau_1$ is equivalent to eliminate the second row (notice that 
we have a simple pole, in this case, i.e. the column 
$( 1 \;\; 0)^{T}$ appears with multiplicity 1).\\
The case $p = 2$ appears in \cite{Brini:2008rh,2008arXiv0804.2592C} and corresponds to a full crepant  
resolution. 
So, by one blow down we arrived at $K_{\mathbb{P}^2(1,1,p)}$
whose charge matrix is given by
\begin{equation}
Q = \left( \begin{array}{cccc} 1\; & \;1\; & \;p \; & \;-p-2 \end{array} \right)
\end{equation}
The associated two sphere partition function is correspondingly
\begin{equation}
\begin{split}
Z=&\,\sum_{m\in\mathbb{Z}}\int \frac{d\tau}{2\pi i} e^{4\pi \xi \tau - i\theta  m} \left(\frac{\Gamma(\tau-\frac{m}{2})}{\Gamma(1-\tau-\frac{m}{2})}\right)^2 \frac{\Gamma(p\tau-p\frac{m}{2})}{\Gamma(1-p\tau-p\frac{m}{2})}\frac{\Gamma(-(p+2)\tau+(p+2)\frac{m}{2}+ irMa)}{\Gamma(1+(p+2)\tau+(p+2)\frac{m}{2}- irMa)} 
\end{split}
\end{equation}
It has two phases, $K_{\mathbb{P}^2(1,1,p)}$ and a more singular $\mathbb{C}^3/\mathbb{Z}_{p+2}$. 
The first phase corresponds to close the integration contour in the left half plane of this effective model; 
since the result is rather ugly, we will simply state that it can be obtained from \eqref{bunwei}, 
with the necessary modifications (i.e. twisted masses). 
For $p=2$ it matches the formula presented in \cite{2008arXiv0804.2592C}.\\

The second phase describing $\mathbb{C}^3/\mathbb{Z}_{p+2}$ can be obtained by flipping the contour to the right half plane and doing the integration in the single variable. Finally, we arrive at
\begin{equation}
\begin{split}
Z &=\, \dfrac{1}{p+2} \sum_{\delta = 0}^{p+1} 
\left( \dfrac{\Gamma(\frac{\delta}{p+2} + \frac{1}{p+2}i r M a )}{\Gamma(1 - \frac{\delta}{p+2} - \frac{1}{p+2}i r M a )} \right)^2 
\dfrac{\Gamma(\langle \frac{p\delta}{p+2}\rangle + \frac{p}{p+2}i r M a )}{\Gamma(1 -\langle \frac{p\delta}{p+2}\rangle - \frac{p}{p+2}i r M a )} \dfrac{1}{(rM)^{2 \left(\delta -\left[ \frac{p \delta}{p+2} \right] \right)}} \\
&\, \sum_{k \geq 0} (-1)^{(p+2) k} (\bar{z}^{-\frac{1}{p+2}})^{(p+2)k + \delta + i r M a} (rM)^{\delta -\left[ \frac{p \delta}{p+2} \right]} \dfrac{(\frac{\delta}{p+2} + \frac{1}{p+2}i r M a )^2_k (\langle \frac{p\delta}{p+2}\rangle + \frac{p}{p+2}i r M a )_{pk+\left[ \frac{p \delta}{p+2} \right]}}{((p+2)k + \delta)!} \\
&\, \sum_{l \geq 0} (-1)^{(p+2) l} (z^{-\frac{1}{p+2}})^{(p+2)l + \delta + i r M a} (-rM)^{\delta -\left[ \frac{p \delta}{p+2} \right]} \dfrac{(\frac{\delta}{p+2} + \frac{1}{p+2}i r M a )^2_l (\langle \frac{p\delta}{p+2}\rangle + \frac{p}{p+2}i r M a )_{pl+\left[ \frac{p \delta}{p+2} \right]}}{((p+2)l + \delta)!} \\
\end{split}
\end{equation}
The ${\cal I}$-function of the orbifold case in the $\delta$-sector of the orbifold cohomology is then obtained from the second line 
of the above formula and for $p=2$ it matches with \cite{2008arXiv0804.2592C}.

\section{Non-abelian GLSM}

In this section we apply our methods to non-abelian gauged linear sigma models and 
give new results for some non-abelian GIT quotients. These are also
tested against results in the mathematical literature when available.

\noindent The first case that we analyse are complex Grassmannians. 
On the way we also give an alternative proof for the conjecture of Hori and Vafa 
which can be rephrased stating that the 
${\cal I}$-function of the Grassmannian can be obtained from that corresponding to a product of 
projective spaces after acting with an appropriate differential operator. 

One can also study a more general theory corresponding to holomorphic vector bundles over Grassmannians. 
These spaces arise in the context of the study of BPS Wilson loop algebra in three dimensional supersymmetric
gauge theories. In particular we will discuss the mathematical counterpart of a duality proposed in 
\cite{2013arXiv1302.2164K} which extends the standard Grassmannian duality to holomorphic vector bundles 
over them.

We also study flag manifolds and more general non-abelian quiver gauge theories for which we provide 
the rules to compute the spherical partition function and the ${\cal I}$-function.

\subsection{Grassmannians}

\noindent The sigma model for the complex Grassmannian $Gr(s,n)$ contains $n$ chirals in the fundamental representation of the $U(s)$ gauge group. Its partition function is given by 
\begin{equation}
Z_{Gr(s,n)} = \dfrac{1}{s!} \sum_{m_1,\ldots , m_s}\int \prod_{i=1}^s \frac{\mathrm{d}\tau_{i}}{2\pi i}e^{4\pi \xi_{\text{ren}} \tau_{i} - i \theta_{\text{ren}} m_i} \prod_{i<j}^s \left( \frac{m_{ij}^2}{4} - \tau_{ij}^2 \right) \prod_{i=1}^s \left( \frac{\Gamma\left(\tau_{i} -\frac{m_i}{2}\right)}{\Gamma\left(1-\tau_{i} -\frac{m_i}{2}\right)}\right)^n 
\end{equation}
As usual, we can write it as
\begin{equation}
\dfrac{1}{s!} \oint \prod_{i=1}^s \dfrac{d (r M\lambda_{i})}{2\pi i} Z_{\text{1l}}Z_{\text{v}}Z_{\text{av}}
\end{equation}
where
\begin{equation}
\begin{split}
Z_{\text{1l}} =&\, \prod_{i=1}^s (r M)^{-2 n r M\lambda_{i}} \left( \frac{\Gamma (rM\lambda_{i})}{\Gamma(1-rM\lambda_{i})}\right)^n  \prod_{i<j}^s (rM\lambda_{i} - rM\lambda_{j})(-rM\lambda_{i} + rM\lambda_{j}) \\
Z_{\text{v}} =&\, z^{-r M \vert \lambda \vert} \sum_{l_1,\ldots , l_s} \dfrac{[(rM)^n (-1)^{s-1} z]^{l_1 + \ldots + l_s}}{(1-rM\lambda_{1})_{l_1}^n \ldots (1-rM\lambda_{s})_{l_s}^n} \prod_{i<j}^s \dfrac{l_i - l_j - rM\lambda_{i} + rM\lambda_{j}}{- rM\lambda_{i} + rM\lambda_{j}} \\
Z_{\text{av}} =&\, \bar{z}^{-r M\vert \lambda \vert} \sum_{k_1,\ldots , k_s} \dfrac{[(-rM)^n (-1)^{s-1} \bar{z}]^{k_1 + \ldots + k_s}}{(1-rM\lambda_{1})_{k_1}^n \ldots (1-rM\lambda_{s})_{k_s}^n} \prod_{i<j}^s \dfrac{k_i - k_j - rM\lambda_{i} + rM\lambda_{j}}{ -rM\lambda_{i}+rM\lambda_{j}}. \\
\end{split}
\end{equation}
We normalized the vortex and antivortex terms in order to have them starting from one in the $rM$ series expansion and we defined $\vert \lambda \vert = \lambda_{1} + \ldots + \lambda_{s}$. The resulting $\mathcal{I}$-function $Z_{\text{v}}$ coincides with the one given in \cite{2003math......4403B}
\begin{equation}
\mathcal{I}_{Gr(s,n)} = e^{\frac{t \sigma_{1}}{\hbar}} \sum_{(d_1, \ldots , d_s)}\dfrac{\hbar^{-n(d_1 + \ldots + d_s)} [(-1)^{s-1}e^t]^{d_1 + \ldots + d_s }}{\prod_{i=1}^s (1+x_i/\hbar)_{d_i}^n } \prod_{i<j}^s \dfrac{d_i - d_j + x_i/\hbar - x_j/\hbar}{x_i/\hbar - x_j/\hbar} 
\end{equation}
if we match the parameters as we did in the previous cases. 
Here the $\lambda$'s are interpreted as Chern roots of the tautological bundle.

\subsubsection{The Hori-Vafa conjecture}

Hori and Vafa conjectured \cite{Hori:2000kt} that $\mathcal{I}_{Gr(s,n)}$ can be obtained by $\mathcal{I}_{\mathbb{P}}$, where $\mathbb{P} = \prod_{i=1}^s \mathbb{P}^{n-1}_{(i)}$, by acting with a differential operator. This has been proved in \cite{2003math......4403B}; here we 
remark that in our formalism this is a simple consequence of the fact that 
the partition function of non-abelian vortices can be obtained from copies of the abelian ones 
upon acting with a suitable differential operator \cite{Bonelli:2011fq}.
In fact we note that $Z_{Gr(s,n)}$ can be obtained from $Z_{\mathbb{P}}$  
simply by dividing by $s!$ and identifying
\begin{equation}
\begin{split}
Z_{\text{1l}}^{Gr} =&\, \prod_{i<j}^s (rM\lambda_{i} - rM\lambda_{j})(-rM\lambda_{i} + rM\lambda_{j}) Z_{\text{1l}}^{\mathbb{P}} \\
Z_{\text{v}}^{Gr}(z) =&\, \prod_{i<j}^s \dfrac{\partial_{z_i}-\partial_{z_j}}{-rM\lambda_{i} + rM\lambda_{j}} Z_{\text{v}}^{\mathbb{P}}(z_1,\ldots , z_s) \Big{\vert}_{z_i=(-1)^{s-1}z} \\
Z_{\text{av}}^{Gr}(\bar{z}) =&\, \prod_{i<j}^s \dfrac{\partial_{\bar{z}_i}-\partial_{\bar{z}_j}}{-rM\lambda_{i} + rM\lambda_{j}} Z_{\text{av}}^{\mathbb{P}}(\bar{z}_1,\ldots , \bar{z}_s) \Big{\vert}_{\bar{z}_i=(-1)^{s-1}\bar{z}}. \\
\end{split}
\end{equation}

\subsection{Holomorphic vector bundles over Grassmannians}
The $U(N)$ gauge theory with $N_f$ fundamentals and $N_a$ antifundamentals flows in the infra-red 
to a non-linear sigma model with target space given by a holomorphic vector bundle of rank $N_a$ 
over the Grassmannian $Gr\left(N,N_{f}\right)$. 
We adopt the notation $Gr\left(N,N_{f}|N_{a}\right)$ for this space. 

One can prove the equality of the partition functions for $Gr\left(N,N_{f}|N_{a}\right)$ and 
$Gr\left(N_{f}-N,N_{f}|N_{a}\right)$ after a precise duality map in a certain range of parameters. 
All this will be specified in the Appendix.
At the level of ${\cal I}$-functions this proves the isomorphism among the 
relevant quantum cohomology rings conjectured in \cite{2013arXiv1302.2164K}.
In analysing this duality we follow the approach of
\cite{Benini:2012ui}, where also the main steps of the proof were outlined. 
However we will detail their calculations and note some differences in the explicit duality map,
which we refine in order to get a precise equality of the partition functions.

The partition function of the 
$Gr\left(N,N_{f}|N_{a}\right)$
GLSM is
\begin{equation}
\begin{split}
Z&=\frac{1}{N!}\sum_{\{m_{s}\in \mathbb{Z}\}_{s=1}^{N}} \int_{(i\mathbb{R})^{N}}\prod_{s=1}^{N}\frac{d\tau_{s}}{2\pi i} z_{\text{ren}}^{-\tau_{s}-\frac{m_{s}}{2}}\bar{z}_{\text{ren}}^{-\tau_{s}+\frac{m_{s}}{2}}\prod_{s<t}^{N}\left( \frac{m_{st}^2}{4} -\tau_{st}^2 \right)\\
& \prod_{s=1}^{N} \prod_{i=1}^{N_{f}}\frac{\Gamma\left(\tau_{s}-i\frac{a_{i}}{\hbar}-\frac{m_{s}}{2}\right)}{\Gamma\left(1-\tau_{s}+i\frac{a_{i}}{\hbar}-\frac{m_{s}}{2}\right)} 
\prod_{s=1}^{N} \prod_{j=1}^{N_{a}}\frac{\Gamma\left(-\tau_{s}+i\frac{\widetilde{a}_{j}}{\hbar}+\frac{m_{s}}{2}\right)}{\Gamma\left(1+\tau_{s}-i\frac{\widetilde{a}_{j}}{\hbar}+\frac{m_{s}}{2}\right)} ,
\end{split}
\end{equation}
while the one of $Gr\left(N_{f}-N,N_{f}|N_{a}\right)$ reads
\begin{equation}
\begin{split}
Z&=\frac{1}{N^{D}!}\sum_{\{m_{s}\in \mathbb{Z}\}_{s=1}^{N^{D}}} \int_{(i\mathbb{R})^{N^{D}}}\prod_{s=1}^{N^{D}}\frac{d\tau_{s}}{2\pi i} (z^{D}_{ren})^{-\tau_{s}-\frac{m_{s}}{2}}(\bar{z}^{D}_{ren})^{-\tau_{s}+\frac{m_{s}}{2}}\prod_{s<t}^{N^{D}}\left(\frac{m_{st}^2}{4} -\tau_{st}^2 \right)\\
& \prod_{s=1}^{N^{D}} \prod_{i=1}^{N_{f}}\frac{\Gamma\left(\tau_{s}+i\frac{a^{D}_{i}}{\hbar}-\frac{m_{s}}{2}\right)}{\Gamma\left(1-\tau_{s}-i\frac{a^{D}_{i}}{\hbar}-\frac{m_{s}}{2}\right)} \prod_{s=1}^{N^{D}} \prod_{j=1}^{N_{a}}\frac{\Gamma\left(-\tau_{s}-i\frac{\widetilde{a}^{D}_{j}}{\hbar}+\frac{m_{s}}{2}\right)}{\Gamma\left(1+\tau_{s}+i\frac{\widetilde{a}^{D}_{j}}{\hbar}+\frac{m_{s}}{2}\right)}\prod_{i=1}^{N_{f}}\prod_{j=1}^{N_{a}}\frac{\Gamma\left(-i\frac{a_{i}-\widetilde{a}_{j}}{\hbar}\right)}{\Gamma\left(1+i\frac{a_{i}-\widetilde{a}_{j}}{\hbar}\right)},
\end{split}
\end{equation}
The proof of the equality of the two is shown in detail in the Appendix to hold
under the duality map
\begin{align}
z^{D}&=(-1)^{N_{a}} z\\
\frac{a^{D}_{j}}{\hbar}&=- \frac{a_{j}}{\hbar}+C\\
\frac{\widetilde{a}^{D}_{j}}{\hbar}&=- \frac{\widetilde{a}_{j}}{\hbar}-(C+i)
\end{align}
where
\begin{equation}
C=\frac{1}{N_{f}-N}\sum_{i=1}^{N_{f}}\frac{a_{i}}{\hbar}.
\end{equation}

\subsection{Flag manifolds}

Let us consider now a linear sigma model with gauge group $U(s_1)\times \ldots \times U(s_l)$ 
and with matter in the $(s_1,\bar{s}_2)\oplus \ldots \oplus (s_{l-1},\bar{s}_l)\oplus (s_l, n)$ 
representations, where $s_1 < \ldots < s_l < n$. This flows in the infrared to a non-linear sigma model 
whose target space is the flag manifold $Fl(s_1,\ldots,s_l,n)$. The partition function is given by
\begin{eqnarray}
Z_{Fl}&=&\dfrac{1}{s_1!\ldots s_l!} \sum_{\substack{\vec{m}^{(a)} \\ a=1\ldots l} }\int \prod_{a=1}^l \prod_{i=1}^{s_a} \frac{\mathrm{d}\tau_{i}^{(a)}}{2\pi i}e^{4\pi \xi^{(a)}_{\text{ren}} \tau_{i}^{(a)} - i \theta^{(a)}_{\text{ren}} m_i^{(a)}} Z_{\text{vector}} Z_{\text{bifund}} Z_{\text{fund}} \nonumber\\
Z_{\text{vector}} &=& \prod_{a=1}^l \prod_{i<j}^{s_a} \left( \frac{(m_{ij}^{(a)})^2}{4} - (\tau_{ij}^{(a)})^2  \right) \nonumber\\
Z_{\text{bifund}} &=& \prod_{a=1}^{l-1} \prod_{i=1}^{s_a}\prod_{j=1}^{s_{a+1}} \frac{\Gamma\left(\tau_{i}^{(a)} - \tau_{j}^{(a+1)} -\dfrac{m_i^{(a)}}{2} + \dfrac{m_j^{(a+1)}}{2}\right)}{\Gamma\left(1-\tau_{i}^{(a)} + \tau_{j}^{(a+1)} -\dfrac{m_i^{(a)}}{2} + \dfrac{m_j^{(a+1)}}{2}\right)} \nonumber\\
Z_{\text{fund}} &=& \prod_{i=1}^{s_l} \left( \frac{\Gamma\left(\tau_{i}^{(l)} -\dfrac{m_i^{(l)}}{2}\right)}{\Gamma\left(1-\tau_{i}^{(l)} -\dfrac{m_i^{(l)}}{2}\right)}\right)^n 
\end{eqnarray}  
This is computed by taking poles at
\begin{equation}
\tau_{i}^{(a)} = \frac{m_i^{(a)}}{2} - k_i^{(a)} + r M \lambda_{i}^{(a)}
\end{equation}
which gives
\begin{eqnarray}
Z_{Fl}&=& \dfrac{1}{s_1!\ldots s_l!} \oint \prod_{a=1}^l \prod_{i=1}^{s_a} \frac{d (r M\lambda_{i}^{(a)})}{2\pi i} Z_{\text{1-loop}}Z_{\text{v}}Z_{\text{av}}
\end{eqnarray}
where
\begin{equation}
\begin{split}
Z_{\text{1-loop}} =& (r M)^{-2rM \left[ \sum_{a=1}^{l-1} (\vert \lambda^{(a)} \vert s_{a+1} - \vert \lambda^{(a+1)} \vert s_{a} ) + n \vert \lambda^{(l)} \vert \right] } \\
& \prod_{a=1}^l \prod_{i<j}^{s_a} (r M\lambda_{i}^{(a)}-r M\lambda_{j}^{(a)})(r M\lambda_{j}^{(a)}-r M\lambda_{i}^{(a)}) \\
&\prod_{a=1}^{l-1} \prod_{i=1}^{s_a}\prod_{j=1}^{s_{a+1}} \frac{\Gamma\left(r M\lambda_{i}^{(a)} - r M\lambda_{j}^{(a+1)} \right)}{\Gamma\left(1-r M\lambda_{i}^{(a)} + r M\lambda_{j}^{(a+1)} \right)} \prod_{i=1}^{s_l} \left( \frac{\Gamma\left(r M\lambda_{i}^{(l)}\right)}{\Gamma\left(1-r M\lambda_{i}^{(l)}\right)}\right)^n \\
Z_{\text{v}} =& \sum_{\vec{l}^{(a)}} (r M)^{ \sum_{a=1}^{l-1} (\vert l^{(a)} \vert s_{a+1} - \vert l^{(a+1)} \vert s_{a} ) + n \vert l^{(l)} \vert } \prod_{a=1}^l (-1)^{(s_a-1) \vert l^{(a)} \vert} z_a^{\vert l^{(a)}\vert - r M \vert \lambda^{(a)} \vert} \\
& \prod_{a=1}^l \prod_{i<j}^{s_a} \dfrac{l_i^{(a)} - l_j^{(a)} - r M \lambda_{i}^{(a)} + r M \lambda_{j}^{(a)}}{-r M \lambda_{i}^{(a)} + r M \lambda_{j}^{(a)}}\\
& \prod_{a=1}^{l-1} \prod_{i=1}^{s_a}\prod_{j=1}^{s_{a+1}} \dfrac{1}{(1-r M\lambda_{i}^{(a)} + r M\lambda_{j}^{(a+1)})_{l^{(a)}_i - l^{(a+1)}_j}} \prod_{i=1}^{s_l} \dfrac{1}{\left[(1-r M\lambda_{i}^{(l)})_{l^{(l)}_i}\right]^n} \\
Z_{\text{av}} =& \sum_{\vec{k}^{(a)}} (-r M)^{\sum_{a=1}^{l-1} (\vert k^{(a)} \vert s_{a+1} - \vert k^{(a+1)} \vert s_{a} ) + n \vert k^{(l)} \vert } \prod_{a=1}^l (-1)^{(s_a-1) \vert k^{(a)} \vert} \bar{z}_a^{\vert k^{(a)} \vert - r M \vert \lambda^{(a)} \vert}\\
& \prod_{a=1}^l \prod_{i<j}^{s_a} \dfrac{k_i^{(a)} - k_j^{(a)} - r M\lambda_{i}^{(a)} + r M\lambda_{j}^{(a)}}{-r M\lambda_{i}^{(a)} + r M\lambda_{j}^{(a)}} \\
& \prod_{a=1}^{l-1} \prod_{i=1}^{s_a}\prod_{j=1}^{s_{a+1}} \dfrac{1}{(1-r M\lambda_{i}^{(a)} + r M\lambda_{j}^{(a+1)})_{k^{(a)}_i - k^{(a+1)}_j}}\prod_{i=1}^{s_l} \dfrac{1}{\left[(1-r M\lambda_{i}^{(l)})_{k^{(l)}_i}\right]^n}\\
\end{split}
\end{equation}
$k$'s and $l$'s are non-negative integers.
\\
This result can be compared with the one in \cite{2004math......7254B}. Indeed
our fractions with Pochhammers at the denominator are equivalent to the products appearing there
and 
we find perfect agreement with the Givental ${\cal I}$-functions under the by now familiar identification 
$\hbar = \frac{1}{r M}, \lambda = - H$ in $Z_{\text{v}}$ and $\hbar = -\frac{1}{r M}, \lambda = H$ 
in $Z_{\text{av}}$.

\subsection{Quivers}

The techniques we used in the flag manifold case can be easily generalized to more general quivers; 
let us write down the rules to compute their partition functions.
Every node of the quiver, i.e. every gauge group $U(s_a)$, contributes with:
\begin{itemize}
\item Integral:
\begin{equation}
\dfrac{1}{s_a !} \oint \prod_{i = 1}^{s_a} \dfrac{d (r M \lambda^{(a)}_i)}{ 2 \pi i} 
\end{equation}
\item One-loop factor:
\begin{equation}
(r M)^{-2 r M \vert \lambda^{(a)} \vert \sum_i Q^{(a)}_i} \prod_{i<j}^{s_a} (r M\lambda_{i}^{(a)}-r M\lambda_{j}^{(a)})(r M\lambda_{j}^{(a)}-r M\lambda_{i}^{(a)})
\end{equation}
\item Vortex factor:
\begin{equation}
\sum_{\vec{l}^{(a)}} (r M)^{ \vert l^{(a)} \vert \sum_i Q^{(a)}_i} (-1)^{(s_a-1) \vert l^{(a)} \vert} z_a^{\vert l^{(a)} \vert - r M \vert \lambda^{(a)} \vert } \prod_{i<j}^{s_a} \dfrac{l_i^{(a)} - l_j^{(a)} - r M \lambda_{i}^{(a)} + r M \lambda_{j}^{(a)}}{-r M \lambda_{i}^{(a)} + r M \lambda_{j}^{(a)}}
\end{equation}
\item Anti-vortex factor:
\begin{equation}
\sum_{\vec{k}^{(a)}} (-r M)^{ \vert k^{(a)}\vert \sum_i Q^{(a)}_i} (-1)^{(s_a-1)\vert k^{(a)} \vert} \bar{z}_a^{\vert k^{(a)} \vert - r M \vert \lambda^{(a)} \vert} \prod_{i<j}^{s_a} \dfrac{k_i^{(a)} - k_j^{(a)} - r M \lambda_{i}^{(a)} + r M \lambda_{j}^{(a)}}{-r M \lambda_{i}^{(a)} + r M \lambda_{j}^{(a)}}
\end{equation}
\end{itemize}
Here $Q^{(a)}_i$ is the charge of the $i$-th chiral matter field with respect to the abelian subgroup $U(1)_{a} \subset U(s_a)$ corresponding to $\xi^{(a)}$ and $ \theta^{(a)}$.\\
\noindent Every matter field in a representation of $U(s_a) \times U(s_b)$ and R-charge $R$ contributes with: 
\begin{itemize}
\item One-loop factor:
\begin{equation}
\prod_{i=1}^{s_a}\prod_{j=1}^{s_{b}} \frac{\Gamma\left(\frac{R}{2} + q_a r M\lambda_{i}^{(a)} + q_b r M\lambda_{j}^{(b)} \right)}{\Gamma\left(1 - \frac{R}{2} - q_a r M\lambda_{i}^{(a)} -q_b r M\lambda_{j}^{(b)} \right)} 
\end{equation}
\item Vortex factor:
\begin{equation}
\prod_{i=1}^{s_a}\prod_{j=1}^{s_{b}} \dfrac{1}{(1 - \frac{R}{2} - q_a r M\lambda_{i}^{(a)} - q_b r M\lambda_{j}^{(b)})_{q_a l^{(a)}_i + q_b l^{(b)}_j}}
\end{equation}
\item Anti-vortex factor:
\begin{equation}
(-1)^{q_a s_b \vert k^{(a)} \vert + q_b s_a \vert k^{(b)} \vert} \prod_{i=1}^{s_a}\prod_{j=1}^{s_{b}} \dfrac{1}{(1 - \frac{R}{2} - q_a r M\lambda_{i}^{(a)} - q_b r M\lambda_{j}^{(b)})_{q_a k^{(a)}_i + q_b k^{(b)}_j}} 
\end{equation}
\end{itemize}
In particular, the bifundamental $(s_a, \bar{s}_b)$ is given by $q_a = 1$, $q_b = -1$. A field in the fundamental can be recovered by setting $q_a = 1$, $q_b = 0$; for an antifundamental, $q_a = -1$ and $q_b = 0$. We can recover the usual formulae if we use \eqref{poc}. Multifundamental representations can be obtained by a straightforward generalization: for example, a trifundamental representation gives 
\begin{equation}
\prod_{i=1}^{s_a}\prod_{j=1}^{s_{b}}\prod_{k=1}^{s_{c}} \dfrac{1}{(1 - \frac{R}{2} - q_a r M\lambda_{i}^{(a)} - q_b r M\lambda_{j}^{(b)} - q_c r M\lambda_{k}^{(c)})_{q_a l^{(a)}_i + q_b l^{(b)}_j + q_c l^{(c)}_k}}
\end{equation}
for the vortex factor.\\
In principle, these formulae are also valid for adjoint fields, if we set $s_a = s_b$, $q_a = 1$, $q_b = -1$; in practice, the diagonal contribution will give a $\Gamma(0)^{s_a}$ divergence, so the only way we can make sense of adjoint fields is by giving them a twisted mass.

\subsection{Orbifold cohomology of the ADHM moduli space}

The formalism described so far has been applied in \cite{Bonelli:2013rja} to the study of 
the equivariant quantum cohomology of the ADHM moduli space.
This is encoded in the following ${\cal I}$-function
\bea
{\cal I}_{k,N}=
\sum_{d_1,\ldots , d_k \,\geq \,0} ((-1)^N z)^{d_1+ \ldots +d_k}  \prod_{r=1}^k \prod_{j=1}^N \dfrac{(-r \lambda_{r}-i r a_{j}+ i r \epsilon)_{d_r}}{(1-r \lambda_{r}-i r a_{j})_{d_r}}
\prod_{r<s}^k \dfrac{d_s - d_r - r \lambda_{s} + r \lambda_{r}}{- r \lambda_{s} + r \lambda_{r}} 
\nonumber
\\
\dfrac{(1+ r \lambda_{r}- r\lambda_{s}-i r \epsilon)_{d_s - d_r}}{( r \lambda_{r}- r\lambda_{s}+i r \epsilon)_{d_s - d_r}} 
\dfrac{( r \lambda_{r}- r\lambda_{s}+i r \epsilon_{1})_{d_s - d_r}}{(1+ r \lambda_{r}- r\lambda_{s}-i r \epsilon_{1})_{d_s - d_r}}
\dfrac{(r \lambda_{r}- r\lambda_{s}+i r \epsilon_{2})_{d_s - d_r}}{(1+ r \lambda_{r}- r\lambda_{s}-i r \epsilon_{2})_{d_s - d_r}}
\nonumber
\\
\label{IKN}\eea
The purpose of this section is to use the wallcrossing approach developed here to
analyze the equivariant quantum cohomology of the Uhlembeck (partial) compactification of the moduli space of
instantons by tuning the FI parameter $\xi$ of the GLSM to zero. Indeed, as we will shortly discuss, 
in this case there is a reflection symmetry $\xi\to - \xi$
showing that the sign of the FI is not relevant to fix the phase of the GLSM.
Actually, fixing $\xi=0$ allows pointlike instantons.
This produces a conjectural formula for the ${\cal I}$-function of the ADHM space in the orbifold chamber.
In particular
for rank one instantons, namely Hilbert schemes of points, our results are in agreement with those in
\cite{2006math.....10129B}.

Let us recall some elementary aspects on the moduli space
${\cal M}_{k,N}$
 of $k$ $SU(N)$ instantons 
on ${\mathbb C}^2$.
This space is non compact both because the manifold ${\mathbb C}^2$ is non compact and because of 
point-like instantons. 
The first source of non compactness is cured by the introduction of the so-called $\Omega$-background which, 
mathematically speaking, corresponds to work in the equivariant cohomology with respect to the maximal torus
of rotations on ${\mathbb C}^2$. The second one can be approached in different ways.
A compactification scheme is provided by the Uhlembeck one
\beq
{\cal M}^U_{k,N}=\bigsqcup_{l=0}^k {\cal M}_{k-l,N} \times S^l\left({\mathbb C}^2\right)
\eeq
Due to the presence of the symmetric product factors this space contains orbifold singularities.
A desingularization is provided by the moduli space of torsion free sheaves on ${\mathbb P}^2$ with a framing on the line at infinity.
This is described in terms of the ADHM complex linear maps
$(B_1,B_2): {\mathbb C}^k\to {\mathbb C}^k$ and 
$(I, J^\dagger): {\mathbb C}^k\to {\mathbb C}^N$ 
which satisfy the F-term equation
$$[B_1,B_2]+IJ=0$$
and the D-term equation
$$[B_1,B_1^\dagger]+[B_2,B_2^\dagger]+II^\dagger - J^\dagger J = \xi {\mathbb I}$$
where $\xi$ is a parameter that gets identified with the FI parameter of the GLSM
and that ensures the stability condition of the sheaf.

Notice that the ADHM equations are symmetric under the reflection 
$\xi\to - \xi$ and 
$$
(B_i,I,J)\to (B_i^\dagger,-J^\dagger,I^\dagger) 
$$
The Uhlembeck compactification is recovered in the $\xi\to 0$ limit.
This amounts to set the vortex expansion parameter as 
\beq
(-1)^Nz=e^{i\theta}
\label{map}
\eeq
giving therefore the orbifold ${\cal I}$-function 
\bea
{\cal I}_{k,N}^U=
\sum_{d_1,\ldots , d_k \,\geq \,0} (e^{i\theta})^{d_1+ \ldots +d_k}  \prod_{r=1}^k \prod_{j=1}^N \dfrac{(-r \lambda_{r}-i r a_{j}+ i r \epsilon)_{d_r}}{(1-r \lambda_{r}-i r a_{j})_{d_r}}
\prod_{r<s}^k \dfrac{d_s - d_r - r \lambda_{s} + r \lambda_{r}}{- r \lambda_{s} + r \lambda_{r}} 
\nonumber
\\
\dfrac{(1+ r \lambda_{r}- r\lambda_{s}-i r \epsilon)_{d_s - d_r}}{( r \lambda_{r}- r\lambda_{s}+i r \epsilon)_{d_s - d_r}} 
\dfrac{( r \lambda_{r}- r\lambda_{s}+i r \epsilon_{1})_{d_s - d_r}}{(1+ r \lambda_{r}- r\lambda_{s}-i r \epsilon_{1})_{d_s - d_r}}
\dfrac{(r \lambda_{r}- r\lambda_{s}+i r \epsilon_{2})_{d_s - d_r}}{(1+ r \lambda_{r}- r\lambda_{s}-i r \epsilon_{2})_{d_s - d_r}}
\nonumber
\\
\label{IKNU}\eea
In the abelian case, namely for $N=1$, the above ${\cal I}$-function reproduces the results of \cite{2006math.....10129B} for the equivariant quantum cohomology of the 
symmetric product of $k$ points in ${\mathbb C}^2$. Indeed, by using the map to the Fock space formalism for the equivariant quantum cohomology
developed in \cite{Bonelli:2013rja}, it is easy to see that both approaches produce the same small equivariant quantum cohomology.
Notice that the map \eqref{map} reproduces in the $N=1$ case the one of \cite{2006math.....10129B}.

\section{$A_p$ and $D_p$ singularities}

The $k$-instanton moduli space for $U(N)$ gauge theories on ALE spaces $\mathbb{C}^2/\Gamma$ has been described by \cite{nakajima} in terms of quiver
representation theory. We can therefore apply the same procedure we used in the previous Section and in \cite{Bonelli:2013rja} and compute the partition function 
on $S^2$ for the relevant quiver. This will give us information about the quantum cohomology of these ALE spaces. Similar results were discussed in \cite{emanuel}.
We will focus on $A_p$ and $D_p$ singularities
and consider the Hilbert scheme of points on their resolutions as well as the orbifold phase given by the symmetric product of points.

Let us start by considering the $A_p$ case.
Define $\vec{k} = (k, \ldots, k)$ vector of $p$ components; the instanton number is given by $k$. The Nakajima quiver describing instantons on $\mathbb{C}^2/\mathbb{Z}_p$ consists of a gauge group $U(k)^p$ with matter $I, J$ in fundamental, antifundamental representation of the first $U(k)$ and matter $B_{b, b \pm 1}$ in bifundamental representations of all the $U(k)$ groups, together with adjoint fields $\chi_b$ and a superpotential $W = \text{Tr}_1[ \chi_{1}( B_{1,2} B_{2,1} - B_{1,p}B_{p,1} + I J )] + \sum_{b=2}^p \text{Tr}_b[ \chi_{b}( B_{b,b+1} B_{b+1,b} - B_{b,b-1}B_{b-1,b} )]$.\footnote{In order to keep a light notation, here $b=p+1$ has to be intended as $b=1$.}

\begin{figure}[h]
\centering
\includegraphics[width=0.3\textwidth]{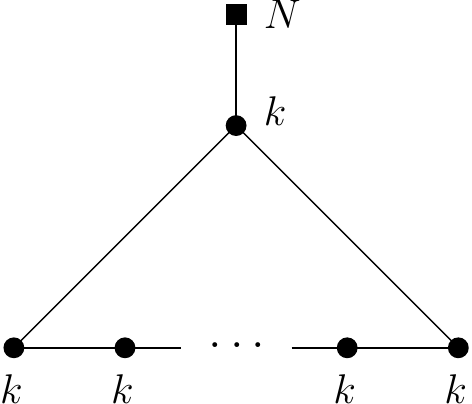}
\caption{The $A_{p-1}^{(1)}$ quiver}
\end{figure}
 
The spherical partition function for this model is given by\footnote{Similarly, here $b=0$ has to be intended as $b=p$.}
\begin{equation}
Z_{\vec{k},N} = \dfrac{1}{(k!)^p} \oint \prod_{b=1}^p \prod_{s=1}^{k} \dfrac{d (r \lambda^{(b)}_s)}{2 \pi i} Z_{\text{1l}} Z_{\text{v}} Z_{\text{av}}   
\end{equation}
\begin{equation}
\begin{split}
Z_{\text{1l}} \,=&\, \left( \dfrac{\Gamma(1- i r \epsilon)}{\Gamma(i r \epsilon)} \right)^{pk}  \prod_{b=1}^p \prod_{s=1}^{k} (z_b \bar{z}_b)^{-r \lambda_s^{(b)}} 
\prod_{b=1}^p \prod_{s=1}^{k} \prod_{t \neq s}^{k} (r \lambda_s^{(b)} - r \lambda_t^{(b)})  \dfrac{\Gamma(1 + r \lambda_s^{(b)} - r \lambda_t^{(b)} - i r \epsilon)}{\Gamma(- r \lambda_s^{(b)} + r \lambda_t^{(b)} + i r \epsilon)}\\
& \prod_{b=1}^p \prod_{s=1}^{k} \prod_{t = 1}^{k} \dfrac{\Gamma(r \lambda_s^{(b)} - r \lambda_t^{(b-1)} + i r \epsilon_1)}{\Gamma(1 - r \lambda_s^{(b)} + r \lambda_t^{(b-1)} - i r \epsilon_1)} \dfrac{\Gamma(- r \lambda_s^{(b)} + r \lambda_t^{(b-1)} + i r \epsilon_2)}{\Gamma(1 + r \lambda_s^{(b)} - r \lambda_t^{(b-1)} - i r \epsilon_2)}\\
& \prod_{s=1}^{k} \prod_{j=1}^{N} \dfrac{\Gamma(r \lambda_s^{(1)} + i r a_j)}{\Gamma(1 - r \lambda_s^{(1)} - i r a_j)} \dfrac{\Gamma(- r \lambda_s^{(1)} - i r a_j + i r \epsilon)}{\Gamma(1 + r \lambda_s^{(1)} + i r a_j - i r \epsilon)}\\
\end{split}
\end{equation}
\begin{equation}
\begin{split}
Z_{\text{v}} \,=&\, \sum_{\{\vec{l}\}}  \prod_{s=1}^{k} (-1)^{N l_s^{(1)}} \prod_{b=1}^p z_b^{l_s^{(b)}} 
\prod_{b=1}^p \prod_{s < t}^{k} \dfrac{l_t^{(b)} - l_s^{(b)} - r \lambda_t^{(b)} + r \lambda_s^{(b)}}{-r \lambda_t^{(b)} + r \lambda_s^{(b)}}  \dfrac{(1 + r \lambda_s^{(b)} - r \lambda_t^{(b)} - i r \epsilon)_{l_t^{(b)} - l_s^{(b)}}}{(r \lambda_s^{(b)} - r \lambda_t^{(b)} + i r \epsilon)_{l_t^{(b)} - l_s^{(b)}}}\\
& \prod_{b=1}^p \prod_{s=1}^{k} \prod_{t = 1}^{k} \dfrac{1}{(1 - r \lambda_s^{(b)} + r \lambda_t^{(b-1)} - i r \epsilon_1)_{l_s^{(b)} - l_t^{(b-1)}}} \dfrac{1}{(1 + r \lambda_s^{(b)} - r \lambda_t^{(b-1)} - i r \epsilon_2)_{l_t^{(b-1)} - l_s^{(b)} }}\\
& \prod_{s=1}^{k} \prod_{j=1}^{N} \dfrac{(- r \lambda_s^{(1)} - i r a_j + i r \epsilon)_{l_s^{(1)}}}{(1 - r \lambda_s^{(1)} - i r a_j)_{l_s^{(1)}}} \\
\end{split}
\end{equation}
\begin{equation}
\begin{split}
Z_{\text{av}} \,=&\, \sum_{\{\vec{k}\}} \prod_{s=1}^{k} (-1)^{N k_s^{(1)}} \prod_{b=1}^p \bar{z}_b^{k_s^{(b)}} 
\prod_{b=1}^p \prod_{s < t}^{k} \dfrac{k_t^{(b)} - k_s^{(b)} - r \lambda_t^{(b)} + r \lambda_s^{(b)}}{-r \lambda_t^{(b)} + r \lambda_s^{(b)}}  \dfrac{(1 + r \lambda_s^{(b)} - r \lambda_t^{(b)} - i r \epsilon)_{k_t^{(b)} - k_s^{(b)}}}{(r \lambda_s^{(b)} - r \lambda_t^{(b)} + i r \epsilon)_{k_t^{(b)} - k_s^{(b)}}}\\
& \prod_{b=1}^p \prod_{s=1}^{k} \prod_{t = 1}^{k} \dfrac{1}{(1 - r \lambda_s^{(b)} + r \lambda_t^{(b-1)} - i r \epsilon_1)_{k_s^{(b)} - k_t^{(b-1)}}} \dfrac{1}{(1 + r \lambda_s^{(b)} - r \lambda_t^{(b-1)} - i r \epsilon_2)_{k_t^{(b-1)} - k_s^{(b)} }}\\
& \prod_{s=1}^{k} \prod_{j=1}^{N} \dfrac{(- r \lambda_s^{(1)} - i r a_j + i r \epsilon)_{k_s^{(1)}}}{(1 - r \lambda_s^{(1)} - i r a_j)_{k_s^{(1)}}} \\
\end{split}
\end{equation}
From $Z_{\text{1l}}$ we can recover in the limit $r \rightarrow 0$ an integral formula for the $A_{p-1}$ ALE Nekrasov partition function:
\begin{equation}
\begin{split}
Z_{\text{ALE}} \,=\, \dfrac{1}{r^{2 N p k}} \dfrac{(i \epsilon)^{p k} }{(k!)^p} \oint & \prod_{b=1}^p \prod_{s=1}^{k} \dfrac{d  \lambda^{(b)}_s}{2 \pi i}
\prod_{b=1}^p \prod_{s=1}^{k} \prod_{t \neq s}^{k} (\lambda_s^{(b)} - \lambda_t^{(b)}) (- \lambda_s^{(b)} + \lambda_t^{(b)} + i \epsilon) \\
& \prod_{b=1}^p \prod_{s=1}^{k} \prod_{t = 1}^{k} \dfrac{1}{(\lambda_s^{(b)} - \lambda_t^{(b-1)} + i \epsilon_1)(- \lambda_s^{(b)} + \lambda_t^{(b-1)} + i \epsilon_2)} \\
& \prod_{s=1}^{k} \prod_{j=1}^{N} \dfrac{1}{(\lambda_s^{(1)} + i a_j)(- \lambda_s^{(1)} - i a_j + i \epsilon)} \\
\end{split}
\end{equation}
We can now study a few examples. In particular, we will be interested in the computation of the equivariant mirror map: this will be non-trivial only in the case $N = 1$, by the same argument proposed in \cite{Bonelli:2013rja}. Even if we are not able to provide a general combinatorial proof, a few examples can convince us that the equivariant mirror map is given by $(1 + \prod_{b=1}^p z_b)^{i k r \epsilon}$, as known from the mathematical literature on the subject \cite{maulik}: 
this has been checked in the cases $k=1, 2$ for $p = 2$ and in the case $k = 1$ for $p = 3, 4$.\\

\noindent We now consider the quiver associated to a $D_{p+1}$ singularity. In this case, the gauge group will be $U(k)^4 \times U(2k)^{p-2}$, with matter $I, J$ in the fundamental, antifundamental representation of the first $U(k)$, matter $B_{b, b\pm 1}$ in bifundamental representations, and matter $\chi_b$ in the adjoint representation, with superpotential 
\bea
W &=& \text{Tr}_1[ \chi_{1}( B_{1,3} B_{3,1} + I J )] + \text{Tr}_2[ \chi_{2}( B_{2,3} B_{3,2} )] 
+ \text{Tr}_3[ \chi_{3}( B_{3,4} B_{4,3} - B_{3,1} B_{1,3} - B_{3,2}B_{2,3} )] \nonumber\\
&&+ \text{Tr}_p[ \chi_{p}( - B_{p,p-1}B_{p-1,p} + B_{p,p+1} B_{p+1,p} + B_{p,p+2} B_{p+2,p} )] + 
\sum_{b=4}^{p-1} \text{Tr}_b[ \chi_{b}( B_{b,b+1} B_{b+1,b} - B_{b,b-1}B_{b-1,b} )] \nonumber\\
&&+ 
\text{Tr}_{p+1}[ \chi_{p+1}( - B_{p+1,p} B_{p,p+1} )] + \text{Tr}_{p+2}[ \chi_{p+2}( - B_{p+2,p} B_{p,p+2} )] \ \ .
\eea
\begin{figure}[h]
\centering
\includegraphics[width=0.5\textwidth]{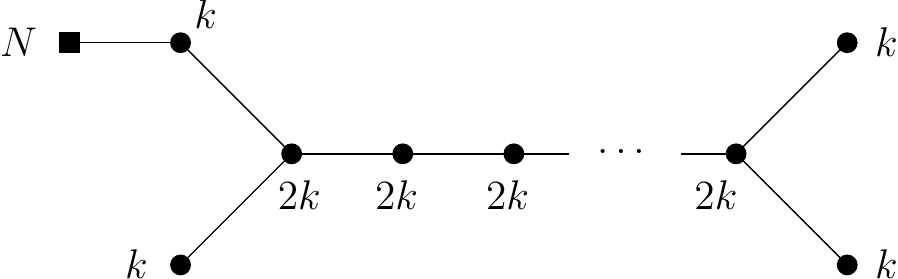}
\caption{The $D_{p+1}^{(1)}$ quiver}
\end{figure}

Defining the $(p+2)$-components vector $\vec{k} = (k,k, 2k, \ldots, 2k, k, k)$, the spherical partition function for this model will be
\begin{equation}
Z_{\vec{k},N} = \dfrac{1}{(k!)^4(2k!)^{p-2}} \oint \prod_{b=1}^{p+2} \prod_{s=1}^{k_b} \dfrac{d (r \lambda^{(b)}_s)}{2 \pi i} Z_{\text{1l}} Z_{\text{v}} Z_{\text{av}}   
\end{equation}
\begin{equation}
\begin{split}
Z_{\text{1l}} \,=&\, \left( \dfrac{\Gamma(1- i r \epsilon)}{\Gamma(i r \epsilon)} \right)^{2 p k}  \prod_{b=1}^{p+2} \prod_{s=1}^{k_b} (z_b \bar{z}_b)^{-r \lambda_s^{(b)}} 
\prod_{b=1}^{p+2} \prod_{s=1}^{k_b} \prod_{t \neq s}^{k_b} (r \lambda_s^{(b)} - r \lambda_t^{(b)})  \dfrac{\Gamma(1 + r \lambda_s^{(b)} - r \lambda_t^{(b)} - i r \epsilon)}{\Gamma(- r \lambda_s^{(b)} + r \lambda_t^{(b)} + i r \epsilon)}\\
& \prod_{b=3}^{p} \prod_{s=1}^{2k} \prod_{t = 1}^{2k} \dfrac{\Gamma(r \lambda_s^{(b+1)} - r \lambda_t^{(b)} + i r \epsilon_1)}{\Gamma(1 - r \lambda_s^{(b+1)} + r \lambda_t^{(b)} - i r \epsilon_1)} \dfrac{\Gamma(- r \lambda_s^{(b+1)} + r \lambda_t^{(b)} + i r \epsilon_2)}{\Gamma(1 + r \lambda_s^{(b+1)} - r \lambda_t^{(b)} - i r \epsilon_2)}\\
& \prod_{b=1}^{2} \prod_{s=1}^{2k} \prod_{t = 1}^{k} \dfrac{\Gamma(r \lambda_s^{(3)} - r \lambda_t^{(b)} + i r \epsilon_1)}{\Gamma(1 - r \lambda_s^{(3)} + r \lambda_t^{(b)} - i r \epsilon_1)} \dfrac{\Gamma(- r \lambda_s^{(3)} + r \lambda_t^{(b)} + i r \epsilon_2)}{\Gamma(1 + r \lambda_s^{(3)} - r \lambda_t^{(b)} - i r \epsilon_2)}\\
& \prod_{b=p+1}^{p+2} \prod_{s=1}^{k} \prod_{t = 1}^{2k} \dfrac{\Gamma(r \lambda_s^{(b)} - r \lambda_t^{(p)} + i r \epsilon_1)}{\Gamma(1 - r \lambda_s^{(b)} + r \lambda_t^{(p)} - i r \epsilon_1)} \dfrac{\Gamma(- r \lambda_s^{(b)} + r \lambda_t^{(p)} + i r \epsilon_2)}{\Gamma(1 + r \lambda_s^{(b)} - r \lambda_t^{(p)} - i r \epsilon_2)}\\
& \prod_{s=1}^{k} \prod_{j=1}^{N} \dfrac{\Gamma(r \lambda_s^{(1)} + i r a_j)}{\Gamma(1 - r \lambda_s^{(1)} - i r a_j)} \dfrac{\Gamma(- r \lambda_s^{(1)} - i r a_j + i r \epsilon)}{\Gamma(1 + r \lambda_s^{(1)} + i r a_j - i r \epsilon)}\\
\end{split}
\end{equation}
\begin{equation}
\begin{split}
Z_{\text{v}} \,=&\, \sum_{\{\vec{l}\}} \prod_{s=1}^{k} (-1)^{N l_s^{(1)}} \prod_{b=1}^{p+2} z_b^{l_s^{(b)}} 
\prod_{b=1}^{p+2} \prod_{s < t}^{k_b} \dfrac{l_t^{(b)} - l_s^{(b)} - r \lambda_t^{(b)} + r \lambda_s^{(b)}}{-r \lambda_t^{(b)} + r \lambda_s^{(b)}}  \dfrac{(1 + r \lambda_s^{(b)} - r \lambda_t^{(b)} - i r \epsilon)_{l_t^{(b)} - l_s^{(b)}}}{(r \lambda_s^{(b)} - r \lambda_t^{(b)} + i r \epsilon)_{l_t^{(b)} - l_s^{(b)}}}\\
& \prod_{b=3}^p \prod_{s=1}^{2k} \prod_{t = 1}^{2k} \dfrac{1}{(1 - r \lambda_s^{(b+1)} + r \lambda_t^{(b)} - i r \epsilon_1)_{l_s^{(b+1)} - l_t^{(b)}}} \dfrac{1}{(1 + r \lambda_s^{(b+1)} - r \lambda_t^{(b)} - i r \epsilon_2)_{l_t^{(b)} - l_s^{(b+1)} }}\\
& \prod_{b=1}^{2} \prod_{s=1}^{2k} \prod_{t = 1}^{k} \dfrac{1}{(1 - r \lambda_s^{(3)} + r \lambda_t^{(b)} - i r \epsilon_1)_{l_s^{(3)} - l_t^{(b)}}} \dfrac{1}{(1 + r \lambda_s^{(3)} - r \lambda_t^{(b)} - i r \epsilon_2)_{l_t^{(b)} - l_s^{(3)} }}\\
& \prod_{b=p+1}^{p+2} \prod_{s=1}^{k} \prod_{t = 1}^{2k} \dfrac{1}{(1 - r \lambda_s^{(b)} + r \lambda_t^{(p)} - i r \epsilon_1)_{l_s^{(b)} - l_t^{(p)}}} \dfrac{1}{(1 + r \lambda_s^{(b)} - r \lambda_t^{(p)} - i r \epsilon_2)_{l_t^{(p)} - l_s^{(b)} }}\\
& \prod_{s=1}^{k} \prod_{j=1}^{N} \dfrac{(- r \lambda_s^{(1)} - i r a_j + i r \epsilon)_{l_s^{(1)}}}{(1 - r \lambda_s^{(1)} - i r a_j)_{l_s^{(1)}}} \\
\end{split}
\end{equation}
\begin{equation}
\begin{split}
Z_{\text{av}} \,=&\, \sum_{\{\vec{k}\}} \prod_{s=1}^{k} (-1)^{N k_s^{(1)}} \prod_{b=1}^{p+2} \bar{z}_b^{k_s^{(b)}} 
\prod_{b=1}^{p+2} \prod_{s < t}^{k_b} \dfrac{k_t^{(b)} - k_s^{(b)} - r \lambda_t^{(b)} + r \lambda_s^{(b)}}{-r \lambda_t^{(b)} + r \lambda_s^{(b)}}  \dfrac{(1 + r \lambda_s^{(b)} - r \lambda_t^{(b)} - i r \epsilon)_{k_t^{(b)} - k_s^{(b)}}}{(r \lambda_s^{(b)} - r \lambda_t^{(b)} + i r \epsilon)_{k_t^{(b)} - k_s^{(b)}}}\\
& \prod_{b=3}^p \prod_{s=1}^{2k} \prod_{t = 1}^{2k} \dfrac{1}{(1 - r \lambda_s^{(b+1)} + r \lambda_t^{(b)} - i r \epsilon_1)_{k_s^{(b+1)} - k_t^{(b)}}} \dfrac{1}{(1 + r \lambda_s^{(b+1)} - r \lambda_t^{(b)} - i r \epsilon_2)_{k_t^{(b)} - k_s^{(b+1)} }}\\
& \prod_{b=1}^{2} \prod_{s=1}^{2k} \prod_{t = 1}^{k} \dfrac{1}{(1 - r \lambda_s^{(3)} + r \lambda_t^{(b)} - i r \epsilon_1)_{k_s^{(3)} - k_t^{(b)}}} \dfrac{1}{(1 + r \lambda_s^{(3)} - r \lambda_t^{(b)} - i r \epsilon_2)_{k_t^{(b)} - k_s^{(3)} }}\\
& \prod_{b=p+1}^{p+2} \prod_{s=1}^{k} \prod_{t = 1}^{2k} \dfrac{1}{(1 - r \lambda_s^{(b)} + r \lambda_t^{(p)} - i r \epsilon_1)_{k_s^{(b)} - k_t^{(p)}}} \dfrac{1}{(1 + r \lambda_s^{(b)} - r \lambda_t^{(p)} - i r \epsilon_2)_{k_t^{(p)} - k_s^{(b)} }}\\
& \prod_{s=1}^{k} \prod_{j=1}^{N} \dfrac{(- r \lambda_s^{(1)} - i r a_j + i r \epsilon)_{k_s^{(1)}}}{(1 - r \lambda_s^{(1)} - i r a_j)_{k_s^{(1)}}} \\
\end{split}
\end{equation}
From $Z_{\text{1l}}$ we can recover an integral expression for the $D_{p+1}$ ALE Nekrasov partition function by taking the limit $r \rightarrow 0$, as we did for the previous case.
The structure of the poles for this model is quite involved, and we leave its study to future work. Nevertheless, an analysis of the simplest cases gives $(1 + z_1 z_2 \prod_{b=3}^p z_b^2 z_{p+1}z_{p+2})^{i r k \epsilon}$ as the equivariant mirror map, again in agreement with \cite{maulik}. 
In line with these computation, we expect also the equivariant mirror map for the $E$-type ALE spaces to depend only on the dual Dynkin label of the affine Dynkin diagram for the corresponding algebra. 

As far as the orbifold phase is concerned, the discussion goes along the same lines as in previous Section 4: by reversing the sign of all Fayet-Iliopoulos
parameters one obtains the same phase due to the symmetry of ADHM constraints. The orbifold phase is then reached by analytic continuation on the product
of circles $|z_b|=1$. This provides conjectural formulae for the equivariant $\mathcal{ I}$ and $\mathcal{J}$ functions of the symmetric product
of points of $A_p$ and $D_p$ singularities that it would be interesting to check against rigorous mathematical results.

\section{Conclusions}
\label{sec:conclusion}

In this paper we exploited some properties of the spherical partition function for supersymmetric $(2,2)$ GLSMs
to provide contour integral formulae for the ${\cal I}$ and the ${\cal J}$-functions encoding the equivariant quantum cohomology of general
GIT quotients. We have given a toolbox to compute the $S^2$ partition function for gauge theory quivers.

We have developed two particular applications of our formulas. 
The first concerns 
the analysis of the contour integral applied to the wall crossing phenomenon among the various chambers of a given GIT quotient.
We used this method to provide conjectural formulae for the quantum cohomology of the ${\mathbb C}^3/{\mathbb Z}_n$ orbifold and 
of the Uhlembeck (partial) compactfication of the instanton moduli space on ${\mathbb C}^2$.
The second has to do with the use of the 
Cauchy theorem to prove gauge theory/quantum cohomology dualities. 
This allowed us to prove a conjectural equivalence of 
quantum cohomology of vector bundles over Grassmannians proposed in the context of the study of Wilson loop algebrae in three dimensional 
supersymmetric gauge theories \cite{2013arXiv1302.2164K}.

There are several directions worth to be further investigated. 
Concerning orbifold quantum cohomology, we underline that our approach can be applied to any classical gauge group and thus
could be exploited for example to compute the Gromov-Witten invariants of $D$ and $E$ type finite groups quotients.

Another interesting issue is the extension of the approach developed in this paper to the computation of open Gromov-Witten invariants
by implementing suitable boundary conditions via the Brini's remodelling technique \cite{2012CMaPh.313..571B}.

Vortex partition functions have been shown to satisfy differential equations of Hypergeometric type
and this has a clear counterpart in the context of AGT correspondence being the null state equations 
for degenerate conformal blocks \cite{Dimofte:2010tz}\cite{Bonelli:2011fq}\cite{Bonelli:2011wx}
\cite{Kozcaz:2010yp}. Differential equations of similar type are obeyed by ${\cal I}$ and ${\cal J}$-functions associated to general
GIT quotients whose explicit form would be useful to spell out in detail in order to study the mirror geometries and the link to classical integrable systems.

These equations are naturally promoted to finite difference equations in K-theoretic vortex counting
\cite{Kozcaz:2010yp}\cite{2001math......8105G}\cite{Bonelli:2011wx}.
The AGT-like dual of these have been recently studied in \cite{Nieri:2013yra}
where their interpretation in terms of q-deformed Virasoro algebra null state equation
is proposed. We plan to study the relation between K-theoretic vortex counting, refined topological strings, quantum K-theory and quantum integrable systems in a forthcoming future.

\section*{Acknowledgments}
We thank F. Benini, 
A. Brini, S. Cremonesi, D.E. Diaconescu, C. Kozcaz, S. Pasquetti and F. Perroni
for interesting discussions and comments.
This research was partly supported by the INFN Research Project PI14 ``Nonperturbative dynamics of gauge theory", 
by the INFN Research Project TV12, 
by   PRIN    ``Geometria delle variet\`a algebriche"
and
by  MIUR-PRIN contract 2009-KHZKRX

\appendix

\section{Duality $Gr\left(N,N_{f}|N_{a}\right) \simeq Gr\left(N_{f}-N,N_{f}|N_{a}\right)$}

\noindent The Grassmannian $Gr\left(N,N_{f}|N_{a}\right)$ is defined as a $U(N)$ gauge theory with $N_{f}$ fundamentals and $N_{a}$ antifundamentals, so we can write the partition function in the form

\begin{equation}
\label{eq:grassmannian partition function}
\begin{split}
Z&=\frac{1}{N!}\sum_{\{m_{s}\in \mathbb{Z}\}_{s=1}^{N}} \int_{(i\mathbb{R})^{N}}\prod_{s=1}^{N}\frac{d\tau_{s}}{2\pi i} z_{\text{ren}}^{-\tau_{s}-\frac{m_{s}}{2}}\bar{z}_{\text{ren}}^{-\tau_{s}+\frac{m_{s}}{2}}\prod_{s<t}^{N}\left( \frac{m_{st}^2}{4} -\tau_{st}^2 \right)\\
& \prod_{s=1}^{N} \prod_{i=1}^{N_{f}}\frac{\Gamma\left(\tau_{s}-i\frac{a_{i}}{\hbar}-\frac{m_{s}}{2}\right)}{\Gamma\left(1-\tau_{s}+i\frac{a_{i}}{\hbar}-\frac{m_{s}}{2}\right)} 
\prod_{s=1}^{N} \prod_{j=1}^{N_{a}}\frac{\Gamma\left(-\tau_{s}+i\frac{\widetilde{a}_{j}}{\hbar}+\frac{m_{s}}{2}\right)}{\Gamma\left(1+\tau_{s}-i\frac{\widetilde{a}_{j}}{\hbar}+\frac{m_{s}}{2}\right)} ,
\end{split}
\end{equation}
where $\hbar$ relates to the radius of the sphere and the renormalization scale $M$ as $\hbar=\frac{1}{r M}$ and $a_{j},\widetilde{a}_{j}$ are the dimensionless (rescaled by $M^{-1}$) equivariant weights for fundamentals and antifundamentals respectively. The renormalized Kahler coordinate $z_{\text{ren}}$ is defined as
\begin{equation}
z_{\text{ren}} = e^{-2\pi \xi_{\text{ren}} + i \theta_{\text{ren}}}= \hbar^{N_{a}-N_{f}}(-1)^{N-1}z.
\end{equation}
since we have
\begin{eqnarray}
\xi_{\text{ren}} = \xi - \frac{1}{2\pi}(N_{f}-N_{a}) \log(r M) \,\,\,,\,\,\, \theta_{\text{ren}} = \theta + (N-1) \pi
\end{eqnarray}
From now on we are setting $M=1$.
We close the contours in the left half planes, so that we pick only poles coming from the fundamentals. We need to build an $N$-pole to saturate the integration measure. Hence the partition function becomes a sum over all possible choices of $N$-poles, i.e. over all combinations how to pick $N$ objects out of $N_{f}$. Now the proposal is that duality holds separately for a fixed choice of an $N$-pole and its corresponding dual. For simplicity of notation let us prove the duality for a particular choice of an $N$-pole and its $(N_{f}-N)$-dual
\begin{equation}
(\underbrace{\Box,\ldots,\Box}_{N},\underbrace{\bullet,\ldots,\bullet}_{N_{f}-N}) \overset{\text{dual}}\longleftrightarrow (\underbrace{\bullet,\ldots,\bullet}_{N},\underbrace{\Box,\ldots,\Box}_{N_{f}-N}),
\end{equation}
where boxes denote the choice of poles forming the $N$-pole.

\subsection{$Gr\left(N,N_{f}|N_{a}\right)$}
The poles are at positions
\begin{equation}
\tau_{s}=-k_{s}+\frac{m_{s}}{2}+\frac{\lambda_{s}}{\hbar}
\end{equation}
and it still remains to be integrated over $\lambda$'s around $\lambda_{s}=i a_{s}$, where $s$ runs from 1 to $N$. This fully specifies from which fundamental we took the pole. Plugging this into~(\ref{eq:grassmannian partition function}), the integral reduces to the following form
\begin{equation}
\label{eq:lambda integral}
\begin{split}
Z=\oint_{\mathcal{M}}\Big\{\prod_{s=1}^{N}\frac{d\lambda_{s}}{2\pi i \hbar}\Big\}
Z_{\text{1l}}\left(\frac{\lambda_{s}}{\hbar},\frac{a_{i}}{\hbar},\frac{\widetilde{a}_{j}}{\hbar}\right)
&z^{-\sum_{s=1}^{N}\frac{\lambda_{s}}{\hbar}}\widetilde{I}\left((-1)^{N_{a}}\kappa z,\frac{\lambda_{s}}{\hbar},\frac{a_{i}}{\hbar},\frac{\widetilde{a}_{j}}{\hbar}\right)\\
\times&\bar{z}^{-\sum_{s=1}^{N}\frac{\lambda_{s}}{\hbar}}\widetilde{I}\left((-1)^{N_{a}}\bar{\kappa}\bar{z},\frac{\lambda_{s}}{\hbar},\frac{a_{i}}{\hbar},\frac{\widetilde{a}_{j}}{\hbar}\right),
\end{split}
\end{equation}
where we defined $\kappa=\hbar^{N_{a}-N_{f}}(-1)^{N-1}$, $\bar{\kappa}=(-\hbar)^{N_{a}-N_{f}}(-1)^{N-1}$. Here we are integrating over a product of circles $\mathcal{M}=\bigotimes_{r=1}^{k}S^{1}(ia_{r},\delta)$ with $\delta$ small enough such that only the pole at the center of the circle is included. From this form we can read of the $I$ function for $Gr\left(N,N_{f}|N_{a}\right)$ as
\begin{equation}
I=z^{-\sum_{s=1}^{N}\frac{\lambda_{s}}{\hbar}}\sum_{\{l_{s}\geq 0\}_{s=1}^{N}}\left((-1)^{N_{a}}\kappa z\right)^{\sum_{s=1}^{N}l_{s}}
\prod_{s<t}^{N}\frac{\lambda_{st}-\hbar  l_{st}}{\lambda_{st}}
\prod_{s=1}^{N}\frac{\prod_{j=1}^{N_{a}}\left(\frac{-\lambda_{s}+i\widetilde{a}_{j}}{\hbar}\right)_{l_{s}}}{\prod_{i=1}^{N_{f}}\left(1+\frac{-\lambda_{s}+ia_{i}}{\hbar}\right)_{l_{s}}},
\end{equation}
where $x_{st}:=x_{s}-x_{t}$.
Now we integrate over $\lambda$'s in~(\ref{eq:lambda integral}), which is straightforward since $Z_{1l}$ contains only simple poles and the rest is holomorphic in $\lambda$'s. Finally, we get
\begin{equation}
Z^{(\Box,\ldots,\Box,\bullet,\ldots,\bullet)}=Z_{\text{class}}Z_{\text{1l}}Z_{\text{v}}Z_{\text{av}},
\end{equation}
where the individual pieces are given as follows
\begin{align}
\label{eq:original theory}
Z_{\text{class}}&=\prod_{s=1}^{N}\left(\hbar^{2(N_{a}-N_{f})}z\bar{z}\right)^{-\frac{ia_{s}}{\hbar}}\\
Z_{\text{1l}}&=\prod_{s=1}^{N}\prod_{i=N+1}^{N_{f}}\frac{\Gamma\left(\frac{ia_{si}}{\hbar}\right)}{\Gamma\left(1-\frac{ia_{si}}{\hbar}\right)} \prod_{s=1}^{N} \prod_{j=1}^{N_{a}}\frac{\Gamma\left(-\frac{i(a_{s}-\widetilde{a}_{j})}{\hbar}\right)}{\Gamma\left(1+\frac{i(a_{s}-\widetilde{a}_{j})}{\hbar}\right)}\\
Z_{\text{v}}&=\sum_{\{l_{s}\geq 0\}_{s=1}^{N}}\left((-1)^{N_{a}}\kappa z\right)^{\sum_{s=1}^{N}l_{s}}
\prod_{s<t}^{N}\left(1-\frac{\hbar  l_{st}}{ia_{st}}\right)
\prod_{s=1}^{N}\frac{\prod_{j=1}^{N_{a}}\left(-i\frac{a_{s}-\widetilde{a}_{j}}{\hbar}\right)_{l_{s}}}{\prod_{i=1}^{N_{f}}\left(1-i\frac{a_{si}}{\hbar}\right)_{l_{s}}}\\
Z_{\text{av}}&=Z_{\text{v}}\left[\kappa z\to \bar{\kappa} \bar{z}\right]
\end{align}
To prove the duality it is actually better to manipulate $Z_{\text{v}}$ to a more convenient form (combining the contributions of the vectors and fundamentals by using identities between the Pochhammers)
\begin{equation}
\label{eq:Zv original}
Z_{\text{v}}=\sum_{l=0}^{\infty}\left[\left(-1\right)^{N_{a} + N - N_{f}}\kappa z\right]^{l}Z_{l}
\end{equation}
with $Z_{l}$ given by
\begin{equation}
\label {eq:original theory vortex}
\begin{split}
Z_{l}=\sum_{\{l_{s}\geq 0|\sum_{s=1}^{N}l_{s}=l\}}\prod_{s=1}^{N}\frac{\prod_{j=1}^{N_{a}}\left(-i\frac{a_{s}-\widetilde{a}_{j}}{\hbar}\right)_{l_{s}}}{l_{s}!\prod_{i\neq s}^{N}\left(i\frac{a_{si}}{\hbar}-l_{s}\right)_{l_{i}}\prod_{i=N+1}^{N_{f}}\left(i\frac{a_{si}}{\hbar}-l_{s}\right)_{l_{s}}}.
\end{split}
\end{equation}

\subsection{The dual theory $Gr\left(N_{f}-N,N_{f}|N_{a}\right)$}
Going to the dual theory not only the rank of the gauge group changes to $N_{f}-N$, but there is a new feature arising. New matter fields $M_{\bar{j}}^{i}$ appear, they are singlets under the gauge group and couple to the fundamentals and antifundamentals via a superpotential $W^{D}=\widetilde{\phi}^{\mu\bar{j}}M_{\bar{j}}^{i}\phi_{\mu i}$. So the partition function gets a new contribution from the mesons $M$ (we set $N^{D}=N_{f}-N$)
\begin{equation}
\begin{split}
Z&=\frac{1}{N^{D}!}\sum_{\{m_{s}\in \mathbb{Z}\}_{s=1}^{N^{D}}} \int_{(i\mathbb{R})^{N^{D}}}\prod_{s=1}^{N^{D}}\frac{d\tau_{s}}{2\pi i} (z^{D}_{ren})^{-\tau_{s}-\frac{m_{s}}{2}}(\bar{z}^{D}_{ren})^{-\tau_{s}+\frac{m_{s}}{2}}\prod_{s<t}^{N^{D}}\left(\frac{m_{st}^2}{4} -\tau_{st}^2 \right)\\
& \prod_{s=1}^{N^{D}} \prod_{i=1}^{N_{f}}\frac{\Gamma\left(\tau_{s}+i\frac{a^{D}_{i}}{\hbar}-\frac{m_{s}}{2}\right)}{\Gamma\left(1-\tau_{s}-i\frac{a^{D}_{i}}{\hbar}-\frac{m_{s}}{2}\right)} \prod_{s=1}^{N^{D}} \prod_{j=1}^{N_{a}}\frac{\Gamma\left(-\tau_{s}-i\frac{\widetilde{a}^{D}_{j}}{\hbar}+\frac{m_{s}}{2}\right)}{\Gamma\left(1+\tau_{s}+i\frac{\widetilde{a}^{D}_{j}}{\hbar}+\frac{m_{s}}{2}\right)}\prod_{i=1}^{N_{f}}\prod_{j=1}^{N_{a}}\frac{\Gamma\left(-i\frac{a_{i}-\widetilde{a}_{j}}{\hbar}\right)}{\Gamma\left(1+i\frac{a_{i}-\widetilde{a}_{j}}{\hbar}\right)},
\end{split}
\end{equation}
where the last factor is the new contribution of the mesons (note that it depends on the original equivariant weights, not on the dual ones). All the computations are analogue to the previous case, so we give the result right after integration
\begin{equation}
Z^{(\bullet,\ldots,\bullet,\Box,\ldots,\Box)}=Z^{D}_{\text{class}}Z^{D}_{\text{1l}}Z^{D}_{\text{v}}Z^{D}_{\text{av}},
\end{equation}
where the building blocks are
\begin{align}
\label{eq:dual theory}
Z^{D}_{\text{class}}&=\prod_{s=N+1}^{N_{f}}\left(\hbar^{2(N_{a}-N_{f})}z^{D}\bar{z}^{D}\right)^{-\frac{ia^{D}_{s}}{\hbar}}\\
Z^{D}_{\text{1l}}&=\prod_{s=N+1}^{N_{f}}\prod_{i=N+1}^{N_{f}}\frac{\Gamma\left(\frac{ia^{D}_{si}}{\hbar}\right)}{\Gamma\left(1-\frac{ia^{D}_{si}}{\hbar}\right)}\prod_{j=1}^{N_{a}}\frac{\Gamma\left(-\frac{i(a^{D}_{s}-\widetilde{a}^{D}_{j})}{\hbar}\right)}{\Gamma\left(1+\frac{i(a^{D}_{s}-\widetilde{a}^{D}_{j})}{\hbar}\right)}\prod_{i=1}^{N_{f}}\prod_{j=1}^{N_{a}}\frac{\Gamma\left(-i\frac{a_{i}-\widetilde{a}_{j}}{\hbar}\right)}{\Gamma\left(1+i\frac{a_{i}-\widetilde{a}_{j}}{\hbar}\right)}\\
Z^{D}_{\text{v}}&=\sum_{l=0}^{\infty}\left[\left(-1\right)^{N_{a}-N}\left(\kappa z\right)^{D}\right]^{l}Z^{D}_{l}\label{eq:Zv dual}\\
Z^{D}_{\text{av}}&=\sum_{k=0}^{\infty}\left[(-1)^{N_{a}-N}\left(\bar{\kappa} \bar{z}\right)^{D}\right]^{k}Z^{D}_{k}
\end{align}
with $Z^{D}_{l}$ given by
\begin{equation}
\label{eq:dual theory vortex}
\begin{split}
Z^{D}_{l}=\sum_{\{l_{s}\geq 0|\sum_{s=N+1}^{N_{f}}l_{s}=l\}}\prod_{s=N+1}^{N_{f}}\frac{\prod_{j=1}^{N_{a}}\left(-i\frac{a^{D}_{s}-\widetilde{a}^{D}_{j}}{\hbar}\right)_{l_{s}}}{l_{s}!\prod_{\substack{i=N+1\\ i\neq s}}^{N_f}\left(i\frac{a^{D}_{si}}{\hbar}-l_{s}\right)_{l_{i}}\prod_{i=1}^{N}\left(i\frac{a^{D}_{si}}{\hbar}-l_{s}\right)_{l_{s}}}.
\end{split}
\end{equation}

\subsection{Duality map}
We are now ready to discuss the duality between the two theories. The statement is the following. For $N_{f}\geq N_{a}+2$, there exists a duality map $z^{D}=z^{D}(z)$ and $a^{D}_{j}=a^{D}_{j}(a_{j}),\;\widetilde{a}^{D}_{j}=\widetilde{a}^{D}_{j}(\widetilde{a}_{j})$ under which the partition functions for $Gr\left(N,N_{f}|N_{a}\right)$ and $Gr\left(N_{f}-N,N_{f}|N_{a}\right)$ are equal.\footnote{We will see the reason for this range later.} In the first step we will construct the duality map and then we will show that~(\ref{eq:original theory}--\ref{eq:original theory vortex}) indeed match with~(\ref{eq:dual theory}--\ref{eq:dual theory vortex}). 
The partition function is a double power series in $z$ and $\bar{z}$ multiplied by $Z_{\text{class}}$. In order to achieve equality of the partition functions, $Z_{\text{class}}$ have to be equal after duality map and then the power series have to match term by term. Moreover we can look only at the holomorphic piece $Z_{\text{v}}$, for the antiholomorphic everything goes in a similar way. The constant term is $Z_{\text{1l}}$, which is a product of gamma functions with arguments linear in the equivariant weights. This implies that the duality map for the equivariant weights is linear. But then the map between the Kahler coordinates can be only a rescaling since a constant term would destroy the matching of $Z_{\text{1l}}$. So we arrive at the most general ansatz for the duality map
\begin{align}
z^{D}&=s z\\
\frac{a^{D}_{i}}{\hbar}&=-E \frac{a_{i}}{\hbar}+C\\
\frac{\widetilde{a}^{D}_{j}}{\hbar}&=-F \frac{\widetilde{a}_{j}}{\hbar}+D
\end{align}
Matching the constant terms $Z_{\text{1l}}$ gives the constraints 
\begin{equation}
E=F=1,\;D=-(C+i).
\end{equation}
Imposing further the equivalence of $Z_{\text{class}}$ fixes $C$ to be
\begin{equation}
C=\frac{1}{N_{f}-N}\sum_{i=1}^{N_{f}}\frac{a_{i}}{\hbar}.
\end{equation}
We are now at a position where $Z_{\text{class}}$ and $Z_{\text{1l}}$ match, while the only remaining free parameter in the duality map is $s$. We fix it by looking at the linear terms in $Z_{\text{v}}$ and $Z^{D}_{\text{v}}$. Of course this does not assure that all higher order terms do match, but we will show that this is the case for $N_{f}\geq N_{a}+2$.\footnote{A direct computation for a handful of examples suggests that higher order terms do not match for $s$ obtained as just outlined if $N_{f}<N_{a}+2$.} So taking only $k=1$ contributions in $Z_{\text{v}}$ and $Z^{D}_{\text{v}}$ we get for $s$
\begin{equation}
s=(-1)^{N-1}\frac{\mathcal{N}}{\mathcal{D}},
\end{equation}
where
\begin{align}
\mathcal{N}&=\sum_{s=1}^{N}\frac{\prod_{j=1}^{N_{a}}\left(-i\frac{a_{s}-\widetilde{a}_{j}}{\hbar}\right)}{\prod_{i\neq s}^{N}\left(-i\frac{a_{si}}{\hbar}\right)\prod_{i=N+1}^{N_{f}}\left(1-i\frac{a_{si}}{\hbar}\right)}\\
\mathcal{D}&=\sum_{s=N+1}^{N_{f}}\frac{\prod_{j=1}^{N_{a}}\left(1+i\frac{a_{s}-\widetilde{a}_{j}}{\hbar}\right)}{\prod_{i=1}^{N}\left(1+i\frac{a_{si}}{\hbar}\right)\prod_{\substack{i=N+1\\j\neq s}}^{N_{f}}\left(-i\frac{a_{si}}{\hbar}\right)}.
\end{align}
The proposal is that for $N_{f}\geq N_{a}+2$
\begin{equation}
s=(-1)^{N_{a}}.
\end{equation}
Out of this range $s$ is a complicated rational function in the equivariant parameters.
This completes the duality map for $N_{f}\geq N_{a}+2$ and suggests that there is no duality map for $N_{f}<N_{a}+2$.

\subsection{Proof of equivalence of the partition functions}
By construction of the mirror map we know that $Z_{\text{class}}$, $Z_{\text{1l}}$ and moreover also the linear terms in $Z_{\text{v}}$  match. Now we will prove (d.m. is the shortcut for duality map)
\begin{equation}
Z_{\text{v}}=Z^{D}_{\text{v}}|_{d.m.}
\end{equation}
for $N_{f}\geq N_{a}+2$. Looking at~\eqref{eq:Zv original} and~\eqref{eq:Zv dual} we see that this boils down to
\begin{equation}
Z_{l}=(-1)^{N_{a}l}Z^{D}_{l}|_{d.m.}.
\end{equation}
The key to prove the above relation is to write $Z_{l}$ as a contour integral
\begin{equation}
\label{eq:vortex contour integral representation}
Z_{l}=\int_{\mathcal{C}_{u}}\prod_{\alpha =1}^{l}\frac{d\phi_{\alpha}}{2\pi i}f\left(\phi,\epsilon,\frac{a}{\hbar},\frac{\widetilde{a}}{\hbar}
\right)\Big|_{\epsilon=1},
\end{equation}
where $\mathcal{C}_{u}$ is a product of contours having the real axes as base and then are closed in the upper half plane by a semicircle. The integrand has the form
\begin{equation}
f=\frac{1}{\epsilon^{l}l!}\prod_{\alpha < \beta}^{l}\frac{\left(\phi_{\alpha}-\phi_{\beta}\right)^{2}}{\left(\phi_{\alpha}-\phi_{\beta}\right)^{2}-\epsilon^{2}}\prod_{\alpha =1}^{l}
\frac{\prod_{j=1}^{N_{a}}\left(i\frac{\widetilde{a}_{j}}{\hbar}+\phi_{\alpha}\right)}{\prod_{i=1}^{N}\left(\phi_{\alpha}+i\frac{a_{i}}{\hbar}\right)\prod_{i=N+1}^{N_{f}}\left(-i\frac{a_{i}}{\hbar}-\epsilon-\phi_{\alpha}\right)}.
\end{equation}
It is necessary to add small imaginary parts to $\epsilon$ and $a_{i}$, $\epsilon\to\epsilon+i\delta,\;-ia_{i}\to-ia_{i}+i\hbar\delta'$ with $\delta>\delta'$.
The proof of~\eqref{eq:vortex contour integral representation} goes by direct evaluation. First we have to classify the poles. Due to the imaginary parts assignments, they are at \footnote[1]{One has to assume $a_i$ to be imaginary at this point. The general result is obtained by analytic continuation after integration.}
\begin{align}
\phi_{\alpha}&=-i\frac{a_{i}}{\hbar} , &&\alpha =1,\ldots, l,\,\,\; i=1,\ldots ,N\\
\phi_{\beta}&=\phi_{\alpha}+\epsilon , && \beta \geq \alpha
\end{align}
We have to build an $l$-pole, which means that the poles are classified by partitions of $l$ into $N$ parts, $l=\sum_{I=1}^{N}l_{I}$. The $I$-th Young tableau $YT(l_{I})$ with $l_{I}$ boxes can be only 1-dimensional (we choose a row) since we have only one $\epsilon$ to play with. To illustrate what we have in mind, we show an example of a possible partition
\begin{equation}
\Yboxdim{10pt}
(\underbrace{\yng(3)}_{l_{1}},\bullet,\yng(2),\yng(1),\ldots,\yng(2),\underbrace{\bullet}_{l_{N}}).
\end{equation}
Residue theorem then turns the integral into a sum over all such partitions and the poles corresponding to a given partition are given as
\begin{equation}
\phi^{I}_{n_{I}}=-i\frac{a_{I}}{\hbar}+(n_{I}-1)\epsilon+\lambda^{I}_{n_{I}},
\end{equation}
where $I=1,\ldots,N$ labels the position of the Young tableau in the $N$-vector and $n_{I}=1,\ldots, l_{I}$ labels the boxes in $YT(l_{I})$.
Substituting this in~\eqref{eq:vortex contour integral representation} we get (the $l!$ gets cancelled by the permutation symmetry of the boxes)
\begin{equation}
\begin{split}
Z_{l}&=
\frac{1}{\epsilon^{l}}\sum_{\{l_{I}\geq 0|\sum_{I=1}^{N}l_{I} = l\}}\oint_{\mathcal{M}}\prod_{\substack{I=1\\l_{I}\neq 0}}^{N}\prod_{n_{I}=1}^{l_{I}}\frac{d\lambda^{I}_{n_{I}}}{2\pi i}\\
&\times\prod_{\substack{I\neq J\\l_{I}\neq 0,l_{J}\neq 0}}^{N}\prod_{n_{I}=1}^{l_{I}}\prod_{n_{J}=1}^{l_{J}}\frac{\left(-i\frac{a_{IJ}}{\hbar}+n_{IJ}\epsilon+\lambda^{I,J}_{n_{I},n_{J}}\right)}{\left(-i\frac{a_{IJ}}{\hbar}+(n_{IJ}-1)\epsilon+\lambda^{I,J}_{n_{I},n_{J}}\right)}
\prod_{\substack{I=1\\l_{I}\neq 0}}^{N}\prod_{n_{I}\neq n_{J}}^{l_{I}}\frac{\left(n_{IJ}\epsilon+\lambda^{I,I}_{n_{I},n_{J}}\right)}{\left((n_{IJ}-1)\epsilon+\lambda^{I,I}_{n_{I},n_{J}}\right)}\\
&\times\prod_{\substack{I=1\\l_{I}\neq 0}}^{N}\prod_{n_{I}=1}^{l_{I}}\frac{\prod_{j=1}^{N_{a}}\left(i\frac{\widetilde{a}_{j}}{\hbar}-i\frac{a_{I}}{\hbar}+(n_{I}-1)\epsilon+\lambda^{I}_{n_{I}}\right)}{\prod_{r=1}^{N}\left(-i\frac{a_{Ir}}{\hbar}+(n_{I}-1)\epsilon+\lambda^{I}_{n_{I}}\right)\prod_{r=N+1}^{N_{f}}\left(-i\frac{a_{Ir}}{\hbar}-n_{I}\epsilon-\lambda^{I}_{n_{I}}\right)},
\end{split}
\end{equation}
where we integrate over $\mathcal{M}=\bigotimes_{r=1}^{l}S^{1}(0,\delta)$.
The computation continues as follows. We separate the poles in $\lambda$'s (there are only simple poles), the rest is a holomorphic function, so we can effectively set the $\lambda$'s to zero there. Eventually, we obtain

\begin{equation}
\begin{split}
Z_{l}&=
\frac{1}{\epsilon^{l}}\sum_{\{l_{I}\geq 0|\sum_{I=1}^{N}l_{I} = l\}} \left[\oint_{\mathcal{M}}\prod_{\substack{I=1\\l_{I}\neq 0}}^{N}\Bigg\{\left(\prod_{n_{I}=1}^{l_{I}}\frac{d\lambda^{I}_{n_{I}}}{2\pi i}\right)\left(\frac{1}{\lambda^{I}_{1}}\prod_{n_{I}=1}^{l_{I}-1}\frac{1}{\lambda^{I,I}_{n_{I}+1,n_{I}}}\right)\Bigg\}\right]\\
&\times\prod_{I\neq J}^{N}\frac{\left(1+i\frac{a_{IJ}}{\hbar\epsilon}-l_{I}\right)_{l_{J}}}{\left(1+i\frac{a_{IJ}}{\hbar\epsilon}\right)_{l_{J}}}\prod_{\substack{I=1\\l_{I}\neq 0}}^{N}\frac{\epsilon^{l_{I}-1}}{l_{I}}\\
&\times\frac{\prod_{I=1}^{N}\prod_{j=1}^{N_{a}}\epsilon^{l_{I}}\left(\frac{i\frac{\widetilde{a}_{j}}{\hbar}+a_{I}}{\epsilon}\right)}{\prod_{I=1}^{N}\prod_{r\neq I}^{N}\epsilon^{l_{I}}\left(-i\frac{a_{Ir}}{\hbar\epsilon}\right)  \prod_{\substack{I=1\\l_{I}\neq 0}}^{N}\epsilon^{l_{I}-1}\left(l_{I}-1\right)! \prod_{I=1}^{N}\prod_{r=N+1}^{N_{f}}\epsilon^{l_{I}}\left(-i\frac{a_{rI}}{\hbar\epsilon}\right)} ,
\end{split}
\end{equation}
where the integration gives $[\ldots]=1$. We are left with products of ratios including the equivariant parameters, which we express as Pochhammer symbols and after heavy Pochhammer algebra we finally arrive at~\eqref{eq:original theory vortex}, which proves~\eqref{eq:vortex contour integral representation}.

\noindent Now, if the integrand ${f}$ does not have poles at infinity, which happens exactly for $N_{f}\geq N_{a}+2$, we can write
\begin{equation}
\int_{\mathcal{C}_{u}}\prod_{\alpha =1}^{l}\frac{d\phi_{\alpha}}{2\pi i}f\left(\phi,\epsilon,\frac{a}{\hbar},\frac{\widetilde{a}}{\hbar}
\right)=(-1)^{l}\int_{\mathcal{C}_{d}}\prod_{\alpha =1}^{l}\frac{d\phi_{\alpha}}{2\pi i}f\left(\phi,\epsilon,\frac{a}{\hbar},\frac{\widetilde{a}}{\hbar}
\right)
\end{equation}
with $\mathcal{C}_{d}$ having the same base as $\mathcal{C}_{u}$ but is closed in the lower half plane by a semicircle. Both contours are oriented counterclockwise. The lovely fact is that the r.h.s. of the above equation gives the desired result
\begin{equation}
(-1)^{l}\int_{\mathcal{C}_{d}}\prod_{\alpha =1}^{l}\frac{d\phi_{\alpha}}{2\pi i}f\left(\phi,\epsilon,\frac{a}{\hbar},\frac{\widetilde{a}}{\hbar}
\right)\Big|_{\epsilon=1}=(-1)^{N_{a}l}Z^{D}_{l}|_{d.m.}
\end{equation}
after direct evaluation of the integral, completely analogue to that of~\eqref{eq:vortex contour integral representation}.

\subsection{Example: the $Gr(1,3) \simeq Gr(2,3)$ case}

Let us show this isomorphism explicitly in a simple case: we will consider $Gr(1,3)$ and $Gr(2,3)$ in a completely equivariant setting.\\
Let us first compute the equivariant partition function for $Gr(1,3)$:
\begin{equation}
\begin{split}
&Z_{Gr(1,3)} = \sum_{m} \int \dfrac{d \tau}{2 \pi i} e^{4 \pi \xi_{\text{ren}} \tau - i \theta_{\text{ren}} m}
\prod_{j = 1}^3 \dfrac{\Gamma(\tau + i r M a_j -\frac{m}{2})}{\Gamma(1 -\tau - i r M a_j -\frac{m}{2})} \\
&\,\,\,= \sum_{i = 1}^3 ((r M)^6 z \bar{z})^{i r M a_i} \prod_{\substack{j = 1\\ j\neq i}}^3 \dfrac{\Gamma(- i r M a_{ij})}{\Gamma(1+i r M a_{ij})} \sum_{l \geq 0} \dfrac{[(r M)^3 z]^l}{\prod_{j=1}^3(1 + i r M a_{ij})_l} \sum_{k \geq 0}  \dfrac{[(-r M)^3 \bar{z}]^k}{\prod_{j=1}^3(1 + i r M a_{ij})_k} \\
\end{split}
\end{equation} 
Here we defined $a_{ij} = a_i-a_j$, and the twisted masses have been rescaled according to $a_i \rightarrow M a_i$, so they are now dimensionless. For $Gr(2,3)$ we have (with $\tilde{\theta}_{\text{ren}} = \tilde{\theta} + \pi = \tilde{\theta} + 3\pi$, being $\tilde{\theta} \longrightarrow \tilde{\theta} + 2 \pi$ a symmetry of the theory)
\begin{equation}
\begin{split}
Z_{Gr(2,3)} =&\, \frac{1}{2}\sum_{m_1, m_2} \int \dfrac{d \tau_{1}}{2 \pi i}\dfrac{d \tau_{2}}{2 \pi i} e^{4 \pi  \tilde{\xi}_{\text{ren}} (\tau_{1} + \tau_{2}) - i \tilde{\theta}_{\text{ren}} (m_1 + m_2)} \\
&\, \left( -\tau_{12}^2 + \frac{m_{12}^2}{4} \right) \prod_{r = 1}^2\prod_{j = 1}^3 \dfrac{\Gamma(\tau_{r} + i r M \tilde{a}_j -\frac{m_{r}}{2})}{\Gamma(1 -\tau_{r} - i r M \tilde{a}_j -\frac{m_{r}}{2})} \\
=&\, \sum_{i<j}^3 ((r M)^6 \tilde{z} \tilde{\bar{z}})^{i r M (\tilde{a}_i+ \tilde{a}_j)} \prod_{\substack{k = 1\\ k\neq i,j}}^3 \dfrac{\Gamma(- i r M \tilde{a}_{ik})}{\Gamma(1+i r M \tilde{a}_{ik})}\dfrac{\Gamma(- i r M \tilde{a}_{jk})}{\Gamma(1+i r M \tilde{a}_{jk})} \\
&\, \sum_{l_1, l_2 \geq 0} \dfrac{[(-r M)^3 \tilde{z}]^{l_1+l_2}}{\prod_{k=1}^3(1 + i r M \tilde{a}_{ik})_{l_1}\prod_{k=1}^3(1 + i r M \tilde{a}_{jk})_{l_2}} \frac{l_1-l_2 +i r M \tilde{a}_i - i r M \tilde{a}_j}{i r M \tilde{a}_i - i r M \tilde{a}_j} \\
&\, \sum_{k_1,k_2 \geq 0}  \dfrac{[(r M)^3 \tilde{\bar{z}}]^{k_1 + k_2}}{\prod_{k=1}^3(1 + i r M \tilde{a}_{ik})_{k_1}\prod_{k=1}^3(1 + i r M \tilde{a}_{jk})_{k_2}}\frac{k_1-k_2 +i r M \tilde{a}_i - i r M \tilde{a}_j}{i r M \tilde{a}_i - i r M \tilde{a}_j} \\
\end{split}
\end{equation} 
In both situations, we are assuming $a_1 + a_2 + a_3 = 0$ and $\tilde{a}_1 + \tilde{a}_2 + \tilde{a}_3 = 0$. Consider now the partition $(\bullet, \bullet, \Box)$ for $Gr(1,3)$ and the dual partition $(\Box, \Box, \bullet)$ for $Gr(2,3)$; we have respectively
\begin{equation}
\begin{split}
Z_{Gr(1,3)}^{(\bullet, \bullet, \Box)} =&\, ((r M)^6 z \bar{z})^{i r M a_3} \dfrac{\Gamma(- i r M a_{31})}{\Gamma(1+i r M a_{31})} \dfrac{\Gamma(- i r M a_{32})}{\Gamma(1+i r M a_{32})} \\
&\, \sum_{l \geq 0} \dfrac{[(r M)^3 z]^l}{l! (1 + i r M a_{31})_l(1 + i r M a_{32})_l} \\
&\, \sum_{k \geq 0}  \dfrac{[(-r M)^3 \bar{z}]^k}{k!(1 + i r M a_{31})_k(1 + i r M a_{32})_k} \\
Z_{Gr(2,3)}^{(\Box, \Box, \bullet)} =&\, ((r M)^6 \tilde{z} \tilde{\bar{z}})^{i r M (\tilde{a}_1+ \tilde{a}_2)} \dfrac{\Gamma(- i r M \tilde{a}_{13})}{\Gamma(1+i r M \tilde{a}_{13})}\dfrac{\Gamma(- i r M \tilde{a}_{23})}{\Gamma(1+i r M \tilde{a}_{23})} \\
&\, \sum_{l_1, l_2 \geq 0} \dfrac{[(-r M)^3 \tilde{z}]^{l_1+l_2}}{\prod_{i=1}^2 l_i! \prod_{j \neq i}^3 (1 + i r M \tilde{a}_{ij})_{l_i}} \frac{l_1-l_2 +i r M \tilde{a}_1 - i r M \tilde{a}_2}{i r M \tilde{a}_1 - i r M \tilde{a}_2}  \\
&\, \sum_{k_1,k_2 \geq 0}  \dfrac{[(r M)^3 \tilde{\bar{z}}]^{k_1 + k_2}}{\prod_{i=1}^2 k_i! \prod_{j \neq i}^3 (1 + i r M \tilde{a}_{ij})_{k_i}}\frac{k_1-k_2 +i r M \tilde{a}_1 - i r M \tilde{a}_2}{i r M \tilde{a}_1 - i r M \tilde{a}_2} \\
\end{split}
\end{equation} 
Since
\begin{equation}
\begin{split}
&\,\sum_{l_1, l_2 \geq 0} \dfrac{[(-r M)^3 \tilde{z}]^{l_1+l_2}}{\prod_{i=1}^2 l_i! \prod_{j \neq i}^3 (1 + i r M \tilde{a}_{ij})_{l_i}} \frac{l_1-l_2 +i r M \tilde{a}_1 - i r M \tilde{a}_2}{i r M \tilde{a}_1 - i r M \tilde{a}_2}\,\, = \\
&=\, \sum_{l \geq 0}\dfrac{[(-r M)^3 \tilde{z}]^{l}}{l!(1 + i r M \tilde{a}_{13})_{l}(1 + i r M \tilde{a}_{23})_{l}} c_l \\
\end{split}
\end{equation} 
and
\begin{equation}
\begin{split}
c_l \,=\, \sum_{l_1 = 0}^l \dfrac{l!}{l_1! (l-l_1)!}\frac{(1 + i r M \tilde{a}_{23} + l - l_1)_{l_1}(1 + i r M \tilde{a}_{13} + l_1)_{l-l_1}}{(i r M \tilde{a}_{12} - l + l_1)_{l_1}(-i r M \tilde{a}_{12} - l_1)_{l-l_1}} \,=\, (-1)^l \,=\, (-1)^{3l}\nonumber
\end{split}
\end{equation} 
we can conclude that $Z_{Gr(1,3)}^{(\bullet, \bullet, \Box)} = Z_{Gr(2,3)}^{(\Box, \Box, \bullet)}$ if we identify $a_i = - \tilde{a}_i$ and $\xi = \tilde{\xi}$, $\theta = \tilde{\theta} $ (i.e., $z = \tilde{z}$). It is then easy to prove that $Z_{Gr(1,3)} = Z_{Gr(2,3)}$.


\providecommand{\href}[2]{#2}\begingroup\raggedright\endgroup

\begin{thebibliography}{10}

\bibitem{2007JHEP...08..052S}
S.~{Shadchin}, {\it {On F-term contribution to effective action}},  {\em
  Journal of High Energy Physics} {\bf 8} (Aug., 2007) 52
  [\href{http://arXiv.org/abs/arXiv:hep-th/0611278}{{\tt
  arXiv:hep-th/0611278}}].

\bibitem{2002hep.th....6161N}
N.~A. {Nekrasov}, {\it {Seiberg-Witten Prepotential From Instanton Counting}},
  {\em ArXiv High Energy Physics - Theory e-prints} (June, 2002)
  [\href{http://arXiv.org/abs/arXiv:hep-th/0206161}{{\tt
  arXiv:hep-th/0206161}}].

\bibitem{Alday:2009aq}
L.~F. Alday, D.~Gaiotto and Y.~Tachikawa, {\it {Liouville Correlation Functions
  from Four-dimensional Gauge Theories}},  {\em Lett.Math.Phys.} {\bf 91}
  (2010) 167--197 [\href{http://arXiv.org/abs/0906.3219}{{\tt 0906.3219}}].

\bibitem{Dimofte:2010tz}
T.~Dimofte, S.~Gukov and L.~Hollands, {\it {Vortex Counting and Lagrangian
  3-manifolds}},  {\em Lett.Math.Phys.} {\bf 98} (2011) 225--287
  [\href{http://arXiv.org/abs/1006.0977}{{\tt 1006.0977}}].

\bibitem{Bonelli:2011fq}
G.~Bonelli, A.~Tanzini and J.~Zhao, {\it {Vertices, Vortices and Interacting
  Surface Operators}},  {\em JHEP} {\bf 1206} (2012) 178
  [\href{http://arXiv.org/abs/1102.0184}{{\tt 1102.0184}}].

\bibitem{Bonelli:2011wx}
G.~Bonelli, A.~Tanzini and J.~Zhao, {\it {The Liouville side of the Vortex}},
  {\em JHEP} {\bf 1109} (2011) 096 [\href{http://arXiv.org/abs/1107.2787}{{\tt
  1107.2787}}].

\bibitem{Kozcaz:2010yp}
C.~Kozcaz, S.~Pasquetti, F.~Passerini and N.~Wyllard, {\it {Affine sl(N)
  conformal blocks from N=2 SU(N) gauge theories}},  {\em JHEP} {\bf 1101}
  (2011) 045 [\href{http://arXiv.org/abs/1008.1412}{{\tt 1008.1412}}].

\bibitem{Kanno:2011fw}
H.~Kanno and Y.~Tachikawa, {\it {Instanton counting with a surface operator and
  the chain-saw quiver}},  {\em JHEP} {\bf 1106} (2011) 119
  [\href{http://arXiv.org/abs/1105.0357}{{\tt 1105.0357}}].

\bibitem{Bulycheva:2012ct}
K.~Bulycheva, H.-Y. Chen, A.~Gorsky and P.~Koroteev, {\it {BPS States in Omega
  Background and Integrability}},  {\em JHEP} {\bf 1210} (2012) 116
  [\href{http://arXiv.org/abs/1207.0460}{{\tt 1207.0460}}].

\bibitem{Benini:2012ui}
F.~Benini and S.~Cremonesi, {\it {Partition functions of N=(2,2) gauge theories
  on $S^2$ and vortices}},  \href{http://arXiv.org/abs/1206.2356}{{\tt
  1206.2356}}.

\bibitem{Doroud:2012xw}
N.~Doroud, J.~Gomis, B.~Le~Floch and S.~Lee, {\it {Exact Results in D=2
  Supersymmetric Gauge Theories}},  {\em JHEP} {\bf 1305} (2013) 093
  [\href{http://arXiv.org/abs/1206.2606}{{\tt 1206.2606}}].

\bibitem{Jockers:2012dk}
H.~Jockers, V.~Kumar, J.~M. Lapan, D.~R. Morrison and M.~Romo, {\it {Two-Sphere
  Partition Functions and Gromov-Witten Invariants}},
  \href{http://arXiv.org/abs/1208.6244}{{\tt 1208.6244}}.

\bibitem{Gomis:2012wy}
J.~Gomis and S.~Lee, {\it {Exact Kahler Potential from Gauge Theory and Mirror
  Symmetry}},  {\em JHEP} {\bf 1304} (2013) 019
  [\href{http://arXiv.org/abs/1210.6022}{{\tt 1210.6022}}].

\bibitem{Park:2012nn}
D.~S. Park and J.~Song, {\it {The Seiberg-Witten Kahler Potential as a
  Two-Sphere Partition Function}},  {\em JHEP} {\bf 1301} (2013) 142
  [\href{http://arXiv.org/abs/1211.0019}{{\tt 1211.0019}}].

\bibitem{Sharpe:2012ji}
E.~Sharpe, {\it {Predictions for Gromov-Witten invariants of noncommutative
  resolutions}},  \href{http://arXiv.org/abs/1212.5322}{{\tt 1212.5322}}.

\bibitem{2013JHEP...05..102H}
Y.~{Honma} and M.~{Manabe}, {\it {Exact K{\"a}hler potential for Calabi-Yau
  fourfolds}},  {\em Journal of High Energy Physics} {\bf 5} (May, 2013) 102
  [\href{http://arXiv.org/abs/1302.3760}{{\tt 1302.3760}}].

\bibitem{Halverson:2013eua}
J.~Halverson, V.~Kumar and D.~R. Morrison, {\it {New Methods for Characterizing
  Phases of 2D Supersymmetric Gauge Theories}},
  \href{http://arXiv.org/abs/1305.3278}{{\tt 1305.3278}}.

\bibitem{Sharpe:2013bwa}
E.~Sharpe, {\it {A few Ricci-flat stacks as phases of exotic GLSM's}},
  \href{http://arXiv.org/abs/1306.5440}{{\tt 1306.5440}}.

\bibitem{1996alg.geom..3021G}
A.~B. {Givental}, {\it {Equivariant Gromov - Witten Invariants}},  in {\em
  eprint arXiv:alg-geom/9603021}, p.~3021, Mar., 1996.

\bibitem{2011arXiv1106.3724C}
I.~{Ciocan-Fontanine}, B.~{Kim} and D.~{Maulik}, {\it {Stable quasimaps to GIT
  quotients}},  {\em ArXiv e-prints} (June, 2011)
  [\href{http://arXiv.org/abs/1106.3724}{{\tt 1106.3724}}].

\bibitem{2013arXiv1302.2164K}
A.~{Kapustin} and B.~{Willett}, {\it {Wilson loops in supersymmetric
  Chern-Simons-matter theories and duality}},  {\em ArXiv e-prints} (Feb.,
  2013) [\href{http://arXiv.org/abs/1302.2164}{{\tt 1302.2164}}].

\bibitem{Witten:1993yc}
E.~Witten, {\it {Phases of N=2 theories in two-dimensions}},  {\em Nucl.Phys.}
  {\bf B403} (1993) 159--222 [\href{http://arXiv.org/abs/hep-th/9301042}{{\tt
  hep-th/9301042}}].

\bibitem{2006math......8481C}
T.~{Coates}, A.~{Corti}, Y.-P. {Lee} and H.-H. {Tseng}, {\it {The Quantum
  Orbifold Cohomology of Weighted Projective Spaces}},  {\em ArXiv Mathematics
  e-prints} (Aug., 2006) [\href{http://arXiv.org/abs/arXiv:math/0608481}{{\tt
  arXiv:math/0608481}}].

\bibitem{2006math.....10129B}
J.~{Bryan} and T.~{Graber}, {\it {The Crepant Resolution Conjecture}},  {\em
  ArXiv Mathematics e-prints} (Oct., 2006)
  [\href{http://arXiv.org/abs/arXiv:math/0610129}{{\tt arXiv:math/0610129}}].

\bibitem{Bershadsky:1993ta}
M.~Bershadsky, S.~Cecotti, H.~Ooguri and C.~Vafa, {\it {Holomorphic anomalies
  in topological field theories}},  {\em Nucl.Phys.} {\bf B405} (1993) 279--304
  [\href{http://arXiv.org/abs/hep-th/9302103}{{\tt hep-th/9302103}}].

\bibitem{Dubrovin:1994hc}
B.~Dubrovin, {\it {Geometry of 2-D topological field theories}},
  \href{http://arXiv.org/abs/hep-th/9407018}{{\tt hep-th/9407018}}.

\bibitem{2001math.....10142C}
T.~{Coates} and A.~{Givental}, {\it {Quantum Riemann - Roch, Lefschetz and
  Serre}},  {\em ArXiv Mathematics e-prints} (Oct., 2001)
  [\href{http://arXiv.org/abs/arXiv:math/0110142}{{\tt arXiv:math/0110142}}].

\bibitem{Bonelli:2013rja}
G.~Bonelli, A.~Sciarappa, A.~Tanzini and P.~Vasko, {\it {The Stringy Instanton
  Partition Function}},  \href{http://arXiv.org/abs/1306.0432}{{\tt
  1306.0432}}.

\bibitem{2006math......3728F}
B.~{Forbes} and M.~{Jinzenji}, {\it {J functions, non-nef toric varieties and
  equivariant local mirror symmetry of curves}},  {\em ArXiv Mathematics
  e-prints} (Mar., 2006) [\href{http://arXiv.org/abs/arXiv:math/0603728}{{\tt
  arXiv:math/0603728}}].

\bibitem{Aganagic:2006wq}
M.~Aganagic, V.~Bouchard and A.~Klemm, {\it {Topological Strings and (Almost)
  Modular Forms}},  {\em Commun.Math.Phys.} {\bf 277} (2008) 771--819
  [\href{http://arXiv.org/abs/hep-th/0607100}{{\tt hep-th/0607100}}].

\bibitem{2007math......2234C}
T.~{Coates}, A.~{Corti}, H.~{Iritani} and H.-H. {Tseng}, {\it {Computing
  Genus-Zero Twisted Gromov-Witten Invariants}},  {\em ArXiv Mathematics
  e-prints} (Feb., 2007) [\href{http://arXiv.org/abs/arXiv:math/0702234}{{\tt
  arXiv:math/0702234}}].

\bibitem{2008arXiv0804.2592C}
T.~{Coates}, {\it {Wall-Crossings in Toric Gromov-Witten Theory II: Local
  Examples}},  {\em ArXiv e-prints} (Apr., 2008)
  [\href{http://arXiv.org/abs/0804.2592}{{\tt 0804.2592}}].

\bibitem{Brini:2008rh}
A.~Brini and A.~Tanzini, {\it {Exact results for topological strings on
  resolved Y**p,q singularities}},  {\em Commun.Math.Phys.} {\bf 289} (2009)
  205--252 [\href{http://arXiv.org/abs/0804.2598}{{\tt 0804.2598}}].

\bibitem{2003math......4403B}
A.~{Bertram}, I.~{Ciocan-Fontanine} and B.~{Kim}, {\it {Two Proofs of a
  Conjecture of Hori and Vafa}},  {\em ArXiv Mathematics e-prints} (Apr., 2003)
  [\href{http://arXiv.org/abs/arXiv:math/0304403}{{\tt arXiv:math/0304403}}].

\bibitem{Hori:2000kt}
K.~Hori and C.~Vafa, {\it {Mirror symmetry}},
  \href{http://arXiv.org/abs/hep-th/0002222}{{\tt hep-th/0002222}}.

\bibitem{2004math......7254B}
A.~{Bertram}, I.~{Ciocan-Fontanine} and B.~{Kim}, {\it {Gromov-Witten
  Invariants for Abelian and Nonabelian Quotients}},  {\em ArXiv Mathematics
  e-prints} (July, 2004) [\href{http://arXiv.org/abs/arXiv:math/0407254}{{\tt
  arXiv:math/0407254}}].

\bibitem{nakajima}
P.B. Kronheimer and H. Nakajima, {\it Yang-Mills instantons on ALE gravitational instantons},
{\em Math.Ann.}{\bf 288} (1990) 263--307.

\bibitem{emanuel}
I. Ciocan-Fontanine, D.-E. Diaconescu and B. Kim, {\it From I to J in two dimensional
$(4,4)$ quiver gauge theories}, preprint.

\bibitem{maulik}
D. Maulik and A. Oblomkov,
    {\it Quantum cohomology of the Hilbert scheme of points on $A_n$-resolutions},
  {\em Journal of the American Mathematical Society} {\bf 22} (2009) 1055-1091,
{\tt arXiv:0802.2737 [math.AG]}.

\bibitem{2012CMaPh.313..571B}
A.~{Brini}, {\it {The Local Gromov-Witten Theory of
  $\{$$\{$C$\}$$\{$P$\}$\^{}1$\}$ and Integrable Hierarchies}},  {\em
  Communications in Mathematical Physics} {\bf 313} (Aug., 2012) 571--605
  [\href{http://arXiv.org/abs/1002.0582}{{\tt 1002.0582}}].

\bibitem{2001math......8105G}
A.~{Givental} and Y.-P. {Lee}, {\it {Quantum K-theory on flag manifolds,
  finite-difference Toda lattices and quantum groups}},  {\em ArXiv Mathematics
  e-prints} (Aug., 2001) [\href{http://arXiv.org/abs/arXiv:math/0108105}{{\tt
  arXiv:math/0108105}}].

\bibitem{Nieri:2013yra}
F.~Nieri, S.~Pasquetti and F.~Passerini, {\it {3d and 5d gauge theory partition
  functions as q-deformed CFT correlators}},
  \href{http://arXiv.org/abs/1303.2626}{{\tt 1303.2626}}.

\end{thebibliography}
  
\end{document}